\documentclass[11pt]{article}
\usepackage{amsmath,amssymb,amsfonts}
\usepackage{cite,graphicx,color}
\usepackage[colorlinks=true, linkcolor=blue, citecolor=blue, linktoc=all]{hyperref}
\usepackage[usenames,dvipsnames]{xcolor}

\begin{document}
\newcommand{\be}{\begin{equation}}
\newcommand{\ee}{\end{equation}}
\newcommand{\beq}{\begin{eqnarray}}
\newcommand{\eeq}{\end{eqnarray}}
\newcommand{\bea}{\begin{eqnarray}}
\newcommand{\eea}{\end{eqnarray}}
\newcommand{\beqn}{\begin{eqnarray}}
\newcommand{\eeqn}{\end{eqnarray}}

\def\pa{\partial}
\newcommand{\un}[1]{\underline{#1}}
\newcommand{\ack}[1]{[{\bf Pfft!: {#1}}]}
\newcommand{\ackm}[1]{\marginnote{\small Pfft!: #1}}
\newcommand{\hlt}[1]{{\color{WildStrawberry}{\em #1}}\index{#1}}

\newcommand{\Dt}{{e}}
\newcommand{\acc}{A}
\newcommand{\LCGamma}{\mathring{\Gamma}}
\newcommand{\LCgamma}{\mathring{\gamma}}
\newcommand{\LCnabla}{\mathring{\nabla}}
\newcommand{\LCtnabla}{\mathring{\tilde{\nabla}}}
\newcommand{\J}{\hat{\mathrm{J}}}
\newcommand{\T}{\hat{\mathrm{T}}}
\newcommand{\Jc}{{\mathrm{J}}}
\newcommand{\Tc}{{\mathrm{T}}}

\newcommand{\thistitle}{
	Weyl Connections and their Role in Holography 
	}
\newcommand{\addresspi}{
	Perimeter Institute for Theoretical Physics, 
	31 Caroline St. N., Waterloo ON, N2L 2Y5, Canada
	}
\newcommand{\addressuiuc}{
	Department of Physics, University of Illinois,
 	1110 West Green St., Urbana IL 61801, U.S.A.
	}
\newcommand{\addresspoly}{
	CPHT, CNRS,
	Ecole Polytechnique, IP Paris,
	91128 Palaiseau Cedex, France
	}
\newcommand{\emailrgl}{rgleigh@illinois.edu}
\newcommand{\emaillc}{ciambelli.luca@gmail.com}
\title{\thistitle}

\author{Luca Ciambelli$^{a,b}$ and Robert G. Leigh$^{b,c}$\\
	\\
	{\small ${}^a$\emph{\addresspoly}}\\ 
	{\small ${}^b$\emph{\addresspi}}\\
	{\small ${}^c$\emph{\addressuiuc}}\\ 
	\\
	}
\date{}
\maketitle
\vspace{-5ex}
\begin{abstract}
\vspace{0.4cm}
It is a well-known property of holographic theories that diffeomorphism invariance in the bulk space-time implies Weyl invariance of the dual holographic field theory in the sense that the field theory couples to a conformal class of background metrics. The usual Fefferman-Graham formalism, which provides us with a holographic dictionary between the two theories, breaks explicitly this symmetry by choosing a specific boundary metric and a corresponding specific metric ansatz in the bulk. In this paper, we show that a simple extension of the Fefferman-Graham formalism allows us to sidestep this explicit breaking; one finds that the geometry of the boundary includes an induced metric and an induced connection on the tangent bundle of the boundary that is a {\it Weyl connection} (rather than the more familiar Levi-Civita connection uniquely determined by the induced metric). Properly invoking this boundary geometry has far-reaching consequences: the holographic dictionary extends and naturally encodes Weyl-covariant geometrical data, and, most importantly, the Weyl anomaly gains a clearer geometrical interpretation, cohomologically relating two Weyl-transformed volumes. The boundary theory is enhanced due to the presence of the Weyl current, which participates with the stress tensor in the boundary Ward identity.
\end{abstract}

\setcounter{footnote}{0}
\renewcommand{\thefootnote}{\arabic{footnote}}
\newpage

\section{Introduction}

The basic principle of general relativity is invariance under diffeomorphisms with, as it is usually formulated, a metric playing the role of the dynamical degrees of freedom. Nonetheless, we usually make use of specific choices of coordinates and parametrizations of the metric, since we are often interested in particular subregions of the space-time manifold. These parametrizations are not harmless in that they break (or gauge fix) some subset of the diffeomorphisms, and one has a restricted class of diffeomorphisms which explicitly preserves the form of a given parametrization. It is most clarifying
to choose a parametrization such that the unbroken symmetries act geometrically on the subregion of spacetime. This is particularly important, for example, for hypersurfaces of any type and co-dimension, but even more generally, for sub-bundles (distributions) of the tangent bundle.

Fefferman and Graham in their seminal works \cite{Feffe, Fefferman2011} found a bulk gauge (FG gauge) preserving the structure of time-like hypersurfaces in AdS$_{d+1}$ spacetimes. This is useful to discuss the time-like conformal boundary, which in suitable coordinates is located at $z=0$, $z$ being the holographic coordinate such that $z=const$ hypersurfaces are time-like. The FG gauge induces on the boundary a metric and its Levi-Civita connection. Although everything is consistent, there exists some leftover freedom in choosing the boundary metric. This comes about because the induced metric on the $z=0$ hypersurface is defined, due to certain bulk diffeomorphisms, up to a rescaling by a non-trivial function of the boundary coordinates. We therefore often refer to the boundary as possessing a conformal class of metrics and say that the boundary enjoys Weyl symmetry. The latter is however often ignored in physical applications, for we usually fix the boundary metric and thus break this symmetry. 

In an attempt to bring electromagnetism and gravity into a unified framework \cite{Weyl1918}, Weyl introduced the concept of Weyl transformation, which encapsulates the possibility of rescaling the metric with an arbitrary scalar function. Weyl symmetry is not considered in many physical systems, but it is a key feature of holography. For instance, it is a very powerful tool in the fluid/gravity correspondence \cite{Bhattacharyya2008a, Loganayagam2008, Bhattacharyya2008b, Ciambelli:2018wre}, where it is exploited in organizing the boundary theory. 

The main observation that we focus on here is that the Levi-Civita connection is not Weyl-covariant, the metricity condition being the source of this non-covariance. This problem can be sidestepped by introducing the notion of a Weyl connection and more generally of Weyl geometry \cite{Folland:1970, Hall:1992}. These concepts have been mentioned in the literature from time to time with reference to a variety of proposed physical applications, mostly in conformal gravitational theory, but also in cosmology and in particle physics, see e.g. \cite{Fulton1962, Matsuo1990, Salim1996, Drechsler1999, Quiros2000, Romero2012, Lobo2015, Mottola2017, Dengiz2017, Cesare2017a, Oda2018, Quiros2018, Ghilencea2019}.\footnote{For a review on applications of Weyl geometry in physics, see \cite{Scholz2018} and references therein.} In the present paper, we will show that Weyl connections play a role in the holographic correspondence, on the field theory side of the duality. Indeed, our first result will be to show that, by slightly generalizing the FG ansatz to what we call the Weyl-Fefferman-Graham gauge (WFG), the {\it Weyl diffeomorphism} responsible for the rescaling of the boundary metric becomes a geometric symmetry. The consequences of this modification are simple: this bulk geometry induces on the boundary a metric and a Weyl connection, instead of its Levi-Civita counterpart. In the dual quantum field theory, these objects act as backgrounds and sources for current operators. Thus, Weyl geometry makes an appearance in holography, not through a modification of the bulk gravitational theory, but in the organization of the dual field theory.

To establish these results, it is important to employ the notion of a (possibly non-integrable) distribution (i.e., a sub-bundle of the tangent bundle), replacing the less general notion of hypersurfaces and foliations. Since this may be unfamiliar to the casual reader, we take some time to review the mathematics, which is informed by theorems of Frobenius. In this way of thinking, the more relevant object is a tangent space, rather than a space itself. 

The FG gauge admits an expansion of the metric from the boundary to the bulk in powers of the holographic coordinate $z$. Solving Einstein equations allows the extraction of the different terms of the expansion, all being determined by two terms in the expansion: the leading order, which defines the boundary conformal class of metrics and the term at order $z^{d-2}$  which  gives the vacuum expectation value of the energy-momentum tensor operator of the dual field theory, as originally discussed in \cite{Maldacena:1997re, Witten:1998qj, Balasubramanian1999d, Skenderis2001}. It is a theorem that, given these two quantities, one can reconstruct, at least order by order, a bulk AdS spacetime in FG gauge --- with some caveats due to the Weyl anomaly, which we will discuss shortly. The resolution of Einstein equations order by order for the WFG gauge on the other hand leads to a modification of the subleading terms in this expansion. In fact, we will demonstrate that the modifications are such that each term is Weyl-covariant; in the FG gauge, the subleading terms transform under Weyl transformations in a very complicated non-linear fashion (which, as we discuss, comes about because they are determined by non-Weyl-covariant Levi-Civita boundary curvature tensors). We will show how to solve Einstein equations in the boundary-to-bulk expansion keeping the space-time dimension $d+1$ arbitrary.

It is a familiar aspect of the FG formalism that the on-shell bulk action diverges as one approaches the boundary. Traditionally, this is dealt with by including local counterterms which are functionals of the induced geometry, in a solution-independent way \cite{Balasubramanian1999d, Emparan1999, Haro2001, Skenderis2002}. There remains one physical subtlety, which is the appearance of a simple pole in $d-2k$, with $k$ integer. This effect is more appropriately thought of as an anomaly in the Weyl Ward identity, a basic feature of renormalization theory \cite{Henningson1998}. This anomaly can be traced back to the fact that holographic renormalization breaks Weyl covariance by fixing a $z=\epsilon$ hypersurface to regulate the theory. No Weyl-covariant renormalization procedures exist. Consequently, a Weyl anomaly is present, and contributes in any even-dimensional boundary theory. There is of course a huge literature on this subject, but an interesting historical account on the Weyl anomaly is \cite{Duff1994}, with a useful list of relevant references therein, as e.g. \cite{Capper1974, Deser1976, Duff1977}. Notice also that a more field-theoretical approach to the anomaly, inspired by string theory and based on the non-invariance of the path integral measure under Weyl transformations can be found in \cite{Polyakov1981, DHoker1988}. The Weyl anomaly is an integral over geometrical tensors, the form of which depends on the dimension. In Ref. \cite{Deser1993},  the geometrical tensors contributing to the anomaly were found in a scheme dependent way. The classification of the Weyl anomaly based on the cohomology of the BRST differential associated with the Weyl symmetry has been performed in \cite{Boulanger:2004zf, Boulanger:2007st, Boulanger:2007ab}. We will unravel a different packaging of the Weyl anomaly, through the use of the WFG gauge ---  the Weyl anomaly will in fact become an integral over Weyl-covariant geometrical tensors. This result reorganizes the theory in a much simpler fashion and opens the door to a  relevant direction of investigation, which is the determination of the coefficients in any even dimension. Indeed, in this relatively short paper, we show how to explicitly construct the anomaly coefficient in three and five bulk dimensions only, while formally deriving its expression in every even boundary dimension, leaving further elaboration to future investigations. We have in fact derived the result in seven and nine dimensions as well, although we do not report the details here, other than to use those results to describe the general structure. Inspired by \cite{Mazur2001}, we will moreover present a simple cohomological interpretation of the Weyl anomaly, based on the difference of two Weyl-related bulk top forms.

The presence of the anomaly is usually encoded in the fact that the boundary energy-momentum tensor acquires an anomalous trace  \cite{Bonora_1986, Osborn1991, Karakhanian1996, DiFrancesco1997}. Indeed in the FG gauge, it is found that it must be {\it a priori} traceless. This boundary Ward identity is obtained by considering the boundary background as dictated by the induced metric only. It is thus natural that there is only one sourced current. However, one finds that one must typically {\it improve} the energy-momentum tensor, as originally found in \cite{Callan1970}. We advocate in this paper a different interpretation, corroborated by the WFG extension. Specifically, we interpret the boundary theory as defined on a background metric (again given by the induced-from-the-bulk metric) and a background Weyl connection, given by the leading order of a bulk dual one-form that occurs in the WFG metric parameterization. In this respect,  two different currents can and indeed do  both participate in the boundary Ward identity. From this perspective we are gauging the Weyl symmetry in the boundary \cite{Iorio1997, Dengiz2012, Coriano2014, Karananas2016}, although more properly, we should view it as a local {\it background} symmetry. Actually, it is the WFG gauge in the bulk that is promoting this Weyl connection to a background configuration in the boundary. We will in particular show that the holographic dictionary furnishes directly this boundary Ward identity relating the trace of the energy-momentum with the divergence of the Weyl current. This will be elegantly verified directly from the boundary action, without invoking holography. Consequently, our setup is useful also to analyze the profound relationship between Weyl invariance and conformal invariance, a subject which has been discussed extensively, for example in \cite{Farnsworth2017, Wu2017} and references therein. While the holographic dictionary will be explored in full detail, the enhancement of the boundary theory will only be briefly described here, with further elaboration left to future works.

The paper is organized as follows. Section \ref{sec2} introduces the Weyl connection, its metricity and torsion properties and its curvature tensors and associated identities. Emphasis is given to its relationship with the ordinary Levi-Civita connection. We then analyze in Section \ref{sec3} the FG gauge and define the Weyl-Fefferman-Graham gauge. We show that the WFG gauge is form-invariant under the Weyl diffeomorphism. We then discuss the important result that we are indeed inducing a Weyl connection on the boundary. The latter makes the (tangent bundle of the) boundary a (generally non-integrable) distribution. Section \ref{sec4} describes the improved holographic dictionary: the boundary Ward identity is derived and it is shown that every term in the bulk-to-boundary expansion is by construction Weyl-covariant. These results are supported by Appendix \ref{AppA}, to which we delegate useful details for the computation of Einstein equations order by order. The next part of this section is devoted to a thorough analysis of the Weyl anomaly, and its cohomological derivation. In this paper, we will confine detailed results to the $d=2$ and $d=4$ cases. Results for $d=6$ and $d=8$ will be reported elsewhere, but some of their key structural aspects will be referred to here.
In Section \ref{sec:QFT}, we present some relevant field theoretical results: we re-derive the Ward identity intrinsically and present examples of simple Weyl-invariant actions. We then conclude and offer some final remarks in Section \ref{conclusion}. 

\section{Weyl Connections and Weyl Manifolds}\label{sec2}

Recall that given a manifold $M$ with metric $g$ and connection $\nabla$ (on the tangent bundle $TM$), we define the metricity $\nabla g$ and torsion $T$ via
\beqn
\nabla_{\un X}g(\un Y,\un Z)&=&\nabla_{\un X}(g(\un Y,\un Z))-g(\nabla_{\un X}\un Y,\un Z)-g(\un Y,\nabla_{\un X}\un Z),\\
T(\un X,\un Y)&=&\nabla_{\un X}\un Y-\nabla_{\un Y}\un X-\left[\un X,\un Y\right],
\eeqn
where $\un{X},...$ are arbitrary vector fields and $\left[\un X,\un Y\right]$ denotes the Lie bracket. Suppose we have a basis $\{\un e_a\}$ of vector fields, and define the connection coefficients via
\beq
\nabla_{\un e_a}\un e_b=\Gamma^c_{ab}\un e_c.
\eeq
It is a familiar theorem that requiring both the metricity and torsion of the connection to vanish leads to a uniquely determined set of connection coefficients, those of the Levi-Civita (LC) connection.  Indeed, further defining the rotation coefficients
\beq
\left[\un e_a,\un e_b\right]=C_{ab}{}^c\un e_c,
\eeq
we find the general result
\beqn
\LCGamma_{ac}^d
= \tfrac{1}{2}g^{db}\Big( \un e_a(g_{bc})+\un e_c(g_{ab})-\un e_b(g_{ca})\Big)-\tfrac{1}{2}g^{db}\Big( C_{ab}{}^fg_{fc}+C_{ca}{}^fg_{fb}-C_{bc}{}^fg_{fa}\Big),
\eeqn
where $g_{ab}\equiv g(\un e_a,\un e_b)$ and we use the circle notation to refer to the LC quantities. This reduces with the choice of coordinate basis $\un e_a=\un\pa_a$ to the familiar Christoffel symbols. 

The vanishing of metricity and torsion are certainly invariant under diffeomorphisms. Therefore, all the geometrical objects built using the LC connection transform nicely under diffeomorphisms. We note though that metricity is not invariant under Weyl transformations\footnote{The Weyl transformation should not be confused with a conformal transformation, which is a diffeomorphism. They do look similar in their actions on the components of the metric,
\beqn
Weyl:&& g_{ab}(x)\mapsto g_{ab}(x)/{\cal B}(x)^2,\\
conformal:&& g_{ab}(x)\mapsto g'_{ab}(x')=g_{ab}(x)/{\omega(x)}^2.
\eeqn
Here though, ${\cal B}(x)$ is an arbitrary function, while $\omega(x)$ is a specific function, associated with a special diffeomorphism that is a conformal isometry.}
 $g\mapsto g/{\cal B}^2$, instead transforming as
\beq 
\nabla g\mapsto (\nabla g-2\text d\ln{\cal B}\otimes g)/{\cal B}^2.
\eeq
Consequently, if we wish to consider geometric theories in which Weyl transformations play a role, it is inconvenient to choose the usual LC connection. Instead, one attains a connection that is covariant with respect to both Weyl transformations and diffeomorphisms by introducing a \hlt{Weyl connection} $A$, \cite{Folland:1970, Hall:1992}, which transforms non-linearly under a Weyl transformation
\beq
g\mapsto g/{\cal B}^2,\qquad A\mapsto A-\text d\ln{\cal B}.
\eeq
By design then, the \hlt{Weyl metricity} is covariant\footnote{To be more specific, what we mean by this notation is 
\[(\nabla g-2A\otimes g)(\un X,\un Y,\un Z)=\nabla_{\un X} g(\un Y,\un Z)-2A(\un X) g(\un Y,\un Z)\]
The notation $A(\un X)$ used here and throughout the paper refers to the contraction of a 1-form with a vector, $A(\un X)\equiv i_{\un X}A\equiv A_aX^a$. }
\beq
(\nabla g-2A\otimes g)\mapsto (\nabla g-2A\otimes g)/{\cal B}^2,
\eeq
and it makes sense to set it to zero if one wishes. Fortunately, there is a theorem which states that there is a unique connection (also generally referred to as a Weyl connection) that has zero torsion and Weyl metricity, see \cite{Hall:1992}. In this case, the connection coefficients are given by the formula
\beqn\label{GammaWeyl}
\Gamma_{ac}^d
 &=& \tfrac{1}{2}g^{db}\Big( \un e_a(g_{bc})+\un e_c(g_{ab})-\un e_b(g_{ca})\Big)-\tfrac{1}{2}g^{db}\Big( C_{ab}{}^fg_{fc}+C_{ca}{}^fg_{fb}-C_{bc}{}^fg_{fa}\Big)\nonumber\\
&&-\Big( A_a \delta^{d}{}_{c}+A_c\delta^{d}{}_{a}-g^{db}A_bg_{ca}\Big).
\eeqn

We note that these connection coefficients are in fact {\it invariant} under Weyl transformations. Consequently, the curvature of the Weyl connection has components\footnote{Here we are using the convention
\beq
{R^a}_{bcd}\un e_a\equiv R(\un e_b,\un e_c,\un e_d)\equiv\nabla_{\un e_c}\nabla_{\un e_d}\un e_b-\nabla_{\un e_d}\nabla_{\un e_c}\un e_b-\nabla_{\left[\un e_c,\un e_d\right]}\un e_b
\eeq
}
\begin{equation}
  \label{eq:riemannchrist}
{R^a}_{bcd}=\un e_c (\Gamma^a_{db})-\un e_d (\Gamma^a_{cb})
+\Gamma^f_{db}\Gamma^a_{cf}-\Gamma^f_{cb}\Gamma^a_{df}-C_{cd}{}^f\Gamma_{fb}^a
\end{equation}
that are themselves Weyl invariant. 
This Weyl-Riemann tensor possesses less symmetries than its Levi-Civita counterpart, and indeed the degrees of freedom contained within are in one-to-one correspondence with the Levi-Civita Riemann tensor, plus a 2-form $F$, which is the field strength $F=\text dA$. To see this, we can write the Weyl curvature components in terms of the 
LC curvature components,
\beqn
R^a{}_{bcd}&=&\mathring{R}^a{}_{bcd}
+\LCnabla_d A_b\delta^a{}_c-\LCnabla_c A_b\delta^a{}_d
+(\LCnabla_d A_c-\LCnabla_c A_d)\delta^a{}_b
+\LCnabla_c A^a g_{bd}-\LCnabla_d A^a g_{bc}
\\&&
+A_b(A_d\delta^a{}_c-A_c\delta^a{}_d)
+A^a(g_{bd}A_c-g_{bc}A_d)
+A^2(g_{bc}\delta^a{}_d-g_{bd}\delta^a{}_c).
\eeqn
The corresponding Weyl-Ricci tensor, which we define as $Ric_{ab}=R^c{}_{acb}$, is given by
\beqn
Ric_{ab}
=\mathring{R}ic_{ab}
-\tfrac{d}{2}F_{ab}
+(d-2)\Big(\LCnabla_{(a} A_{b)}+A_a A_b\Big)
+\Big(\LCnabla\cdot A-(d-2)A^2\Big)g_{ab}
\eeqn
in space-time dimension $d$.
We then read off that the Weyl-Ricci tensor has an antisymmetric part
\beq
Ric_{[ab]}=-\tfrac{d}{2}F_{ab},
\eeq
while the symmetric part differs from the LC Ricci tensor,
\beqn
Ric_{(ab)}
=\mathring{R}ic_{ab}
+(d-2)\Big(\LCnabla_{(a} A_{b)}+A_a A_b\Big)
+\Big(\LCnabla\cdot A-(d-2)A^2\Big)g_{ab}.
\eeqn
The corresponding Weyl-Ricci scalar is the trace,
\beq
R
=\mathring{R}
+2(d-1)\LCnabla\cdot A-(d-1)(d-2)A^2.
\eeq
Under a Weyl transformation, $R\to R{\cal B}^2$, so we see that the LC Ricci scalar must transform very non-trivially under Weyl,
\beq
\mathring{R}\mapsto {\cal B}^2\Big(\mathring{R}+2(d-1)\mathring\nabla^2\ln{\cal B}
-(d-1)(d-2)(\pa\ln{\cal B})^2\Big)
\eeq
in order to cancel the transformation of the non-Weyl-invariant expression involving the Weyl connection $A$. Similarly, $Ric_{ab}\mapsto Ric_{ab}$ implies
\beq
\mathring{R}ic_{ab}\mapsto\mathring{R}ic_{ab}+g_{ab}\mathring{\nabla}\cdot\pa\ln{\cal B}+(d-2)\Big(\mathring{\nabla}_{(a}\pa_{b)}\ln{\cal B}+\pa_{a}\ln{\cal B}\pa_{b}\ln{\cal B}-g_{ab}(\pa\ln{\cal B})^2\Big)
\eeq
We thus see the important role played by the Weyl connection. Organizing the theory with respect to the latter is a more natural prescription, whenever this theory includes Weyl transformations.

Given a Weyl connection, we can organize tensors in such a way that they have a specific Weyl weight and we use the notation
\beq\label{WeylGauged}
\hat\nabla_{\un X} t=\nabla_{\un X} t+w_t A(\un X)\ t.
\eeq
whereby
\beq
t\mapsto {\cal B}^{w_t}t,\qquad \hat\nabla t\mapsto {\cal B}^{w_t}\hat\nabla t.
\eeq
For the specific case of a scalar field $\phi$, we would then write $\hat\nabla_a\phi=\un e_a(\phi)+w_\phi A_a\phi$. The condition that Weyl metricity vanishes is translated in this notation as $\hat\nabla g=0$.

Finally we remark that the Bianchi identity for the Weyl-Riemann tensor is
\beq\label{fullBianchi}
\nabla_a R^e{}_{bcd}+\nabla_c R^e{}_{bda}+\nabla_d R^e{}_{bac}=0
\eeq
Contracting the $e,c$ indices, we get the once-contracted Bianchi identity
\beq
\nabla_a Ric_{bd}-\nabla_d Ric_{ba}+\nabla_c R^c{}_{bda}=0.
\eeq
which given that the Weyl-Riemann and Weyl-Ricci tensors are Weyl invariant (that is, they have weight zero), can also be written as 
\beq\label{ocBianchi}
\hat\nabla_a Ric_{bd}-\hat\nabla_d Ric_{ba}+\hat\nabla_c R^c{}_{bda}=0.
\eeq

If we multiply by $g^{ab}$, we find
\beq
g^{ab}\hat\nabla_a Ric_{bd}-\hat\nabla_d R+\hat\nabla_c(g^{ab} R^c{}_{bda})=0.
\eeq
This can be simplified further by noting that
\beq
g^{ab} R^c{}_{bda}=g^{cb}\Big(Ric_{bd}+2F_{bd}\Big)
\eeq
and hence the twice contracted Bianchi identity can be simplified to
\beq
g^{ab}\hat\nabla_a (G_{bc}+F_{bc})=0
\eeq
where $G_{ab}=Ric_{ab}-\tfrac12 Rg_{ab}$ is the Weyl-Einstein tensor. Since $G$ and $F$ have Weyl weight zero, this can also be written as
\beq\label{tcBianchi}
g^{ab}\nabla_a (G_{bc}+F_{bc})=0
\eeq
This is the analogue of the familiar result in Riemannian geometry, $\LCnabla^a \mathring{G}_{ac}=0$.

\section{Weyl Invariance and Holography}\label{sec3}

The Fefferman-Graham theorem \cite{Feffe, Fefferman2011} says that the metric of a locally asymptotically AdS$_{d+1}$ (LaAdS) geometry can always be put in the form
\beq
\text ds^2=L^2\frac{\text dz^2}{z^2}+h_{\mu\nu}(z;x)\text dx^\mu\text dx^\nu.
\eeq
The conformal boundary is a constant-$z$ hypersurface at $z=0$ in these coordinates. To obtain this form, one has used up all of the diffeomorphism invariance, apart from residual transformations of the $x^\mu\mapsto x'{}^\mu(x)$, which of course would change the components of $h_{\mu\nu}$ in general. 

Near $z=0$, $h_{\mu\nu}(z;x)$ may be expanded
\beq
\label{FGexpandmetricBetter}
h_{\mu\nu}(z;x)=\frac{L^2}{z^2}\left[\gamma^{(0)}_{\mu\nu}(x)+\frac{z^2}{L^2}\gamma^{(2)}_{\mu\nu}(x)+\frac{z^4}{L^4}\gamma^{(4)}_{\mu\nu}(x)+...\right]+\frac{z^{d-2}}{L^{d-2}}\left[\pi^{(0)}_{\mu\nu}(x)+\frac{z^2}{L^2}\pi^{(2)}_{\mu\nu}(x)+...\right]+....
\eeq
Here, we are regarding the boundary dimension $d$ as variable\footnote{This holographic dimensional regularization avoids the necessary introduction of logarithms (as in e.g. \cite{Skenderis2002}) that occur when $d$ is an even integer.} (in fact, we will regard $d\in\mathbb{C}$ formally as needed. This is discussed further later in the paper.) Given this expansion, $\gamma^{(0)}_{\mu\nu}(x)$ has an interpretation as an induced boundary metric:
\beqn\label{conf}
\dfrac{z^2}{L^2}\text ds^2 &\underset{z\to 0}{\longrightarrow}&\gamma^{(0)}_{\mu\nu}(x)\text dx^\mu \text dx^\nu=\text ds^2_{bdy}.
\eeqn
It is this object that sources the stress energy tensor in the dual field theory, with $\pi^{(0)}_{\mu\nu}(x)$ its vev. All of the other terms in the expansion are determined in terms of $\gamma^{(0)}_{\mu\nu}(x),\pi^{(0)}_{\mu\nu}(x)$ by the bulk classical equations of motion. 

Equation \eqref{conf} defines the induced boundary metric up to a Weyl transformation. We see indeed that there is an ambiguity in the construction of this metric which amounts in defining the latter up to a scalar function of the boundary coordinates. Although it is often stated, this ambiguity is usually disregarded and a specific choice is made. 

The following bulk diffeomorphism (which we refer to as the \hlt{Weyl diffeomorphism})
\beq\label{Weyldiff}
z\mapsto z'=z/{\cal B}(x),\qquad x^\mu\mapsto x'{}^\mu=x^\mu
\eeq
plays an important role. It has the effect of inducing a Weyl transformation of the boundary metric components: using \eqref{conf} with  holographic coordinate now $z'$ we obtain
\beq
\text ds^2_{bdy}=\frac{\gamma^{(0)}_{\mu\nu}(x)}{{\cal B}(x)^2}\text dx^\mu \text dx^\nu.
\eeq
However, this diffeomorphism does not leave the bulk metric in the Fefferman-Graham gauge, but rather transforms it to
\beq\label{shiftedmetric}
\text ds^2= L^2\left(\frac{\text dz'}{z'}+\pa_\mu\ln{\cal B}(x)\ \text dx^\mu\right)^2
+h_{\mu\nu}(z'{\cal B}(x);x)\text dx^\mu \text dx^\nu
\eeq
where
\beqn\label{Weyltransmetric}
h_{\mu\nu}(z'{\cal B}(x);x)&=&\frac{L^2}{z'^2}\left[\frac{\gamma^{(0)}_{\mu\nu}(x)}{{\cal B}(x)^2}+\frac{z'^2}{L^2}\gamma^{(2)}_{\mu\nu}(x)+\frac{z'^4}{L^4}{\cal B}(x)^2\gamma^{(4)}_{\mu\nu}(x)+...\right]\\
&&\qquad+\frac{z'^{d-2}}{L^{d-2}}\left[{\cal B}(x)^{d-2}\pi^{(0)}_{\mu\nu}(x)+\frac{z'^2}{L^2}{\cal B}(x)^d\pi^{(2)}_{\mu\nu}(x)+...\right]+....
\eeqn
Thus, this diffeomorphism takes us out of FG gauge (as it is one of the diffs that was fixed in going to that gauge), and acts on the boundary tensors $\gamma_{\mu\nu}^{(k)}(x)$ and $\pi_{\mu\nu}^{(k)}(x)$ by a local Weyl rescaling with specific $k$-dependent weights. 

The standard way to deal with the fact that we have been taken out of FG gauge is to employ an additional diffeomorphism acting on the $x^\mu\mapsto x^\mu+\xi^\mu(z;x)$ which becomes trivial at the conformal boundary in such a way that $\gamma^{(0)}_{\mu\nu}(x)$ is left unchanged, but the cross term in (\ref{shiftedmetric}) is cancelled (see e.g. \cite{Mazur2001}). However, this diffeomorphism unfortunately has a complicated effect on all of the subleading terms in the metric --- they no longer transform linearly as in (\ref{Weyltransmetric}), but instead transform non-linearly under the combined transformations and, we claim, this obscures the geometric significance of the sub-leading terms. There is nothing inconsistent here: in FG gauge, the subleading terms are given on-shell by expressions involving the Levi-Civita curvature of the induced metric, which themselves transform non-linearly under Weyl transformations. 

The fact that the Weyl diffeomorphism has taken us out of the Fefferman-Graham gauge, and in particular acts on a piece of the bulk metric other than $h_{\mu\nu}$ as in eq. \eqref{shiftedmetric}, motivates replacing the Fefferman-Graham gauge by
\beq\label{WFGmetric}
\text ds^2=L^{2}\left(\frac{\text dz}{z}-a_\mu(z;x) \text dx^\mu\right)^2
+h_{\mu\nu}(z;x)\text dx^\mu \text dx^\nu.
\eeq
which we refer to as Weyl-Fefferman-Graham (WFG) gauge. In this form, the bulk metric is given in terms of two tensor fields, $h_{\mu\nu}$ and $a_\mu$, and the Weyl diffeomorphism acts as
\beq\label{shiftedmetricW}
\text ds^2= L^2\left(\frac{\text dz'}{z'}-a_\mu(z'{\cal B}(x);x)\text dx^\mu+\pa_\mu\ln{\cal B}(x)\ \text dx^\mu\right)^2
+h_{\mu\nu}(z'{\cal B}(x);x)\text dx^\mu \text dx^\nu.
\eeq
Thus, the Weyl diffeomorphism can be interpreted as acting on the fields $h_{\mu\nu}$ and $a_\mu$, preserving the form of the WFG gauge without the need for a compensating diffeomorphism; the action on $h_{\mu\nu}$ is as before while $a_\mu$ shifts non-linearly and so ultimately will be interpreted as a connection or gauge field. We will be precise about the details of these transformations below.

In WFG gauge, the constant-$z$ hypersurface $\Sigma$ at $z=0$ remains the conformal boundary with induced metric $\gamma^{(0)}$, as
\beqn\label{conf2}
\dfrac{z^2}{L^2}\text ds^2 &\underset{z\to 0}{\longrightarrow}&\gamma^{(0)}_{\mu\nu}(x)\text dx^\mu \text dx^\nu.
\eeqn
Thus the presence of $a_\mu$ in the ansatz does not modify the induced metric at $z=0$. 
However, as we will see, this does not mean that $a_\mu$ does not appear at the conformal boundary. This is surprising since $a_\mu$ is pure gauge in the bulk, but we will see that  $a_\mu$ has a clear geometric interpretation in the boundary theory. To understand this claim, we first note that the metric is no longer diagonal in the $z,x^\mu$ coordinates, and so we must take greater care in interpreting how we approach the conformal boundary. We now describe that process carefully.

It is natural, given the metric ansatz (\ref{WFGmetric}), to introduce the 1-form
\beq\label{defefordistrib}
e\equiv \Omega(z;x)^{-1}\left(\frac{\text dz}{z}-a_\mu(z;x) \text dx^\mu\right)
\eeq
This form defines a \hlt{distribution} $C_e\subset TM$ defined as
\beq
C_e=ker(e)=span\left\{\un X\in \Gamma(TM)\Big| i_{\un X}e=0\right\}.
\eeq
Note that there is an ambiguity in multiplying $e$ (or equivalently the $\un X$'s) by a function on $M$, and we have represented this ambiguity by introducing the function $\Omega$ in \eqref{defefordistrib}. 

We remark that if $a_\mu$ were zero, then $C_e$ would be the span of the vectors  $\un\pa_\mu$, which form a basis for the tangent spaces to constant-$z$ hypersurfaces. In the present context though, this more general  notion of a distribution is the appropriate geometrical structure. In the general case, it is
convenient to introduce a basis for $C_e$ as the set of vectors
\beq
\un D_\mu\equiv\un\pa_\mu+a_\mu(z;x)z\un\pa_z.
\eeq 
This implies that we can regard $a_\mu$ as providing a lift\footnote{Here, we are regarding $\Sigma$ as an isolated hypersurface in $M$. We can thus regard $M$ as a fibre bundle $\pi:M\to \Sigma$. An Ehresmann connection provides a splitting of the tangent bundle $TM=H\oplus V$, and the $\un D_\mu$ vectors form a basis of $H$, identified with $C_e$, at the point $(z,x^\mu)$. \label{ftn:ehresmann}} from $T\Sigma$ (with basis $\{\un\pa_\mu\}$) to $C_e$, that is, it can be thought of as an Ehresmann connection. 
By the Frobenius theorem, $C_e$ is an {\it integrable} distribution if 
\beq
\left[\un D_\mu,\un D_\nu\right] \in C_e.
\eeq
To understand this condition, it is convenient to introduce a vector dual to $e$,
\beq\label{standarde}
\un e\equiv \Omega(z;x)z\un\pa_z
\eeq
which has been normalized to $i_{\un e}e=e(\un e)=1$, and we regard $\{\un e,\un D_\mu\}$ as a basis for $T_{(z;x)}M$. We then compute
\beq\label{commutatorsonC}
\left[\un D_\mu,\un D_\nu\right]= \Omega(z;x)^{-1}f_{\mu\nu}(z;x)\un e,\qquad f_{\mu\nu}(z;x)\equiv D_\mu a_\nu(z;x)-D_\nu a_\mu(z;x).
\eeq
So we find that integrability is the condition $f_{\mu\nu}=0$, and thus by Frobenius, the distribution $C_e$ would define under that circumstance a foliation of $M$ by co-dimension one hypersurfaces. We will not find it necessary to assume that the distribution is in fact integrable, and thus we will not assume $a_\mu$ to be flat.

By taking $\un e$ in the form (\ref{standarde}), we have fixed some of the diffeomorphism invariance;\footnote{Indeed, the vector field $\un e$ could more generally be of the form 
\beq
\un e\to \un e'=\un e+\theta^\mu(z;x)\un D_\mu
\eeq
which satisfies $e(\un e)=1$ for any $\theta^\mu$. (In the language of footnote \ref{ftn:ehresmann} 
the $\un e$ of (\ref{standarde}) is special in that $\un e\in V$.) In the general case, we have $\left[\un D_\mu,\un D_\nu\right]= f_{\mu\nu}\un e'-f_{\mu\nu}\theta^\lambda\un D_\lambda$ and thus integrability remains the condition $f_{\mu\nu}=0$. The second diffeomorphism, discussed earlier, that returns the metric to the FG ansatz after a boundary Weyl transformation corresponds on the contrary to setting $a_\mu\to 0$ at the expense of keeping $\theta^\mu\neq 0$.
} the residual diffeomorphisms that preserve the form of $\un e$ in \eqref{standarde} are given by $z'=z'(z;x), x'{}^\mu=x'{}^\mu(x)$.This set of diffeomorphisms includes but is larger than the Weyl diffeomorphisms. Given the interpretation of holography in terms of renormalization, we expect that these diffeomorphisms correspond to generic {\it local} (in $x$) coarse grainings. This is ultimately the reason for our construction, which retains a clear covariant geometric interpretation for this more general notion of renormalization.\footnote{In most field theory contexts, such local rescalings are not considered. However, it is clear that they are of general interest. For example, it is widely appreciated that ultraviolet divergences that occur in calculations of entanglement observables arise through the pile-up of modes near the entanglement cut, and renormalization towards the cut is the natural procedure. We should also mention that similar structure is known to arise in holographic fluids.}
These residual diffeomorphisms act on the form $e$ as
\beq\label{defefordistribtrans}
e\mapsto e'=\Omega'(z'(z;x);x'(x))^{-1}\left(\frac{\text dz'(z;x)}{z'(z;x)}-a'_\mu(z'(z;x);x'(x)) \text dx'{}^\mu\right)
\eeq
and thus leave it invariant if
\beq\label{ressymmtrans}
\frac{\pa x'{}^\nu(x)}{\pa x{}^\mu}a'_\nu(z';x') = \frac{\pa \ln z'(z;x)}{\pa\ln z}a_\mu(z;x)
+\frac{\pa\ln z'(z;x)}{\pa x{}^\mu},\qquad 
\Omega'(z';x')=\frac{\pa\ln z'(z;x)}{\pa\ln z}\Omega(z;x).
\eeq
The first equation is consistent with the interpretation of $a$ as an Ehresmann connection.
The second equation implies that the inherent ambiguity in the definition of the distribution $C_e$ represented by $\Omega(z;x)$ can be thought of as the (local) reparametrization invariance of $z$. We can for example use this reparametrization invariance to set $\Omega(z;x)\to L^{-1}$ if we wish. The diffeomorphisms that preserve this choice (or, more generally preserve any specific $\Omega(z;x)$) are a subset of the aforementioned residual diffeomorphisms and are of the form $z'=z/{\cal B}(x), x'{}^\mu=x'{}^\mu(x)$, which are precisely the Weyl diffeomorphisms (together with an arbitrary reparameterization $x'=x'(x)$) that preserve the form of the metric \eqref{WFGmetric}. In this case, the first equation in \eqref{ressymmtrans} reduces to
\beq
\frac{\pa x'{}^\nu(x)}{\pa x{}^\mu}a'_\nu(z';x') = a_\mu(z;x)
-\pa_\mu\ln {\cal B}(x),
\eeq
and so we are to interpret the $a_\mu(z;x)$ as a connection for the Weyl diffeomorphisms (\ref{Weyldiff}). This transformation differs from what appears in eq. \eqref{shiftedmetricW} only because here we are allowing for an arbitrary transverse diffeomorphism  $x'=x'(x)$ as well. Given this result, it may not come as a surprise that $a_\mu(z;x)$ will induce a Weyl connection on the conformal boundary, and we will establish precisely that below. 

To recap, we have been led to the following (non-coordinate) basis for the tangent space
\beq
\{\un e,\un D_\mu\}= \Big\lbrace{L^{-1}z\un\pa_z,\un\pa_\mu+a_\mu z\un\pa_z \Big\rbrace}
\eeq
which have the following Lie brackets
\beq
\left[\un D_\mu,\un D_\nu\right]= Lf_{\mu\nu}\un e,\qquad
\left[\un D_\mu,\un e\right]= -L\un e(a_\mu) \un e
\eeq
To proceed further, we Fourier analyze $a_\mu(z;x)$ and $h_{\mu\nu}(z;x)$ in the sense that we will expand them in eigenfunctions of $\un e$. Such eigenfunctions are of course just the monomials in $z\in \mathbb{R}^+$. For $h_{\mu\nu}(z;x)$ we obtain then the same expansion as before, eq. (\ref{FGexpandmetricBetter}), and for 
 $a_\mu(z;x)$ we write
\beq\label{WFGexpandgauge}
a_{\mu}(z;x)=\left[a^{(0)}_{\mu}(x)+\frac{z^2}{L^2}a^{(2)}_{\mu}(x)+...\right]
+\frac{z^{d-2}}{L^{d-2}}\left[p^{(0)}_{\mu}(x)+\frac{z^2}{L^2}p^{(2)}_{\mu}(x)+...\right]+...,
\eeq
which is of the same form as the expansion of a massless gauge field in Fefferman-Graham gauge \cite{Witten:1998qj}. One may have expected that since $a_\mu$ is a pure gauge part of the bulk metric, it should not source a (vector) operator in the boundary theory. However, what we will show is that  $a^{(0)}_\mu$ is not part of the boundary metric but will appear instead as part of the induced boundary connection. It thus represents a choice that we should have of selecting a connection which is not the Levi-Civita connection determined entirely by a choice of the induced boundary metric. 

More precisely, what we will show is that for the WFG ansatz, the induced connection is {\it not} the Levi-Civita connection of the induced metric, but instead a Weyl connection. 
Given the expansions (\ref{FGexpandmetricBetter},\ref{WFGexpandgauge}), we see that the Weyl diffeomorphism (\ref{Weyldiff}) acts as
\beqn\label{Weyldifftrans}
\gamma^{(k)}_{\mu\nu}(x)\mapsto\gamma^{(k)}_{\mu\nu}(x){\cal B}(x)^{k-2},\qquad
\pi^{(k)}_{\mu\nu}(x)\mapsto \pi^{(k)}_{\mu\nu}(x){\cal B}(x)^{d-2+k}\\
a^{(k)}_{\mu}(x)\mapsto a^{(k)}_{\mu}(x){\cal B}(x)^{k}-\delta_{k,0}\pa_\mu\ln{\cal B}(x),\qquad
p^{(k)}_{\mu}(x)\mapsto p^{(k)}_{\mu}(x){\cal B}(x)^{d-2+k}\label{Weyldifftrans2}
\eeqn
and so in particular
\beq
\gamma^{(0)}_{\mu\nu}(x)\mapsto\gamma^{(0)}_{\mu\nu}(x)/{\cal B}(x)^{2},\qquad
a^{(0)}_{\mu}(x)\mapsto a^{(0)}_{\mu}(x)-\pa_\mu\ln{\cal B}(x)
\eeq
and thus we may anticipate that $a^{(0)}_\mu$ will play the role of a boundary Weyl connection. All of the other subleading functions in the expansions (\ref{FGexpandmetricBetter},\ref{WFGexpandgauge}) are interpreted to have, \`a la (\ref{Weyldifftrans}--\ref{Weyldifftrans2}), definite Weyl weights, that is they are Weyl tensors. It is then natural to expect that they will be determined in terms of the Weyl curvature, which we discussed in the last section.

We introduced the concept of the distribution $C_e$ precisely in order to properly discuss the notion of an induced connection, as $C_e$ is a sub-bundle of $TM$.  That is, given a bulk connection $\nabla$ on $TM$ (which we will take to be the LC connection), we can apply it to vectors in $C_e$, which will be of the general form
\beq\label{definducedconn}
\nabla_{\un D_\mu}\un D_\nu=\Gamma_{\mu\nu}^\lambda \un D_\lambda+\Gamma_{\mu\nu}^e\un e.
\eeq
The coefficients of the induced connection on $C_e$ are by definition the $\Gamma_{\mu\nu}^\lambda$ appearing in (\ref{definducedconn}). Notice that these connection coefficients should not be confused with the usual Christoffel symbols, which are associated with coordinate bases.  By direct computation, we find
\beqn
\Gamma_{\mu\nu}^\lambda=\gamma_{\mu\nu}^\lambda\equiv
\tfrac12h^{\lambda\rho}\Big( D_\mu h_{\rho\nu} +D_\nu h_{\mu\rho} - D_\rho h_{\nu\mu}\Big)
\eeqn
and furthermore if we evaluate this expression at $z=0$, we find
\beq\label{InducedWeyl}
\gamma^{(0)}{}_{\mu\nu}^\lambda=\tfrac12\gamma_{(0)}^{\lambda\rho}\Big(
 (\pa_\mu -2 a^{(0)}_\mu) \gamma^{(0)}_{\nu\rho}
 +(\pa_\nu -2 a^{(0)}_\nu) \gamma^{(0)}_{\mu\rho}
 -(\pa_\rho -2 a^{(0)}_\rho) \gamma^{(0)}_{\mu\nu}\Big).
\eeq
This result can be compared to (\ref{GammaWeyl}), from which  we conclude that the induced connection on the boundary is in fact a Weyl connection, with the role of the  geometric data $g_{ab}$ and $A_a$ in \eqref{GammaWeyl} being played here by $\gamma^{(0)}_{\mu\nu}$ and $a^{(0)}_\mu$. In comparing, we make use of the fact that here the intrinsic rotation coefficients are $C_{\mu\nu}{}^\lambda=0$, as in (\ref{commutatorsonC}). We will use the notation $\nabla^{(0)}$ for the corresponding Weyl connection (whose Weyl-Christoffel symbols are given by \eqref{InducedWeyl}), and the curvature as $R^{(0)}{}^\lambda{}_{\mu\rho\nu}$. A tensor with components $t_{\mu_1...\mu_n}(x)$ that has Weyl weight $w_t$ transforms as $t_{\mu_1...\mu_n}(x)\mapsto {\cal B}(x)^{w_t}t_{\mu_1...\mu_n}(x)$, while $\hat\nabla^{(0)}_\nu t_{\mu_1...\mu_n}(x)\equiv \nabla^{(0)}_\nu t_{\mu_1...\mu_n}(x)+w_ta^{(0)}_\nu t_{\mu_1...\mu_n}(x)$ transforms covariantly with the same weight. As noted above, all of the component fields aside from $a^{(0)}_\mu$ transform covariantly with respect to arbitrary Weyl transformations, and the Weyl weights of the various component fields are given above in \eqref{Weyldifftrans}.
In the next section, we will briefly study some aspects of the holographic dictionary, and we will find that every equation is covariant with respect to arbitrary Weyl transformations --- it is a {\it bona fide} (background) symmetry of the dual field theory. 
In particular, we will find that the appearance of $a^{(0)}_\mu(x)$, since it transforms non-linearly under Weyl transformations, is through Weyl-covariant derivatives of other fields, or through expressions involving the Weyl-invariant field strength $f^{(0)}_{\mu\nu}$. Before moving on, we would like to stress again the main result of this section: the usual bulk LC connection built using the bulk metric in the enhanced WFG gauge induces on the boundary a Weyl connection and therefore a boundary Weyl-covariant geometry.

\section{The Holographic Dictionary and the Weyl Anomaly}\label{sec4}

In this section, we will explore some details of the holographic dictionary corresponding to the WFG ansatz. The LC connection in the bulk has the form
\beqn\label{conn1}
\nabla_{\un D_\mu}\un D_\nu&=&
\gamma_{\mu\nu}^\lambda\un D_\lambda
-h_{\nu\lambda}\psi^\lambda{}_\mu\un e
\\\label{conn2}
\nabla_{\un D_\mu}\un e&=&\psi^\lambda{}_\mu\un D_\lambda
\\\label{conn3}
\nabla_{\un e}\un D_\mu&=&\psi^\lambda{}_\mu\un D_\lambda +L\varphi_\mu\un e
\\\label{conn4}
\nabla_{\un e}\un e&=&-Lh^{\lambda\rho}\varphi_\rho \un D_\lambda
\eeqn
where 
\beq\label{connComp}
\psi^\mu{}_\nu=\rho^\mu{}_\nu+\frac{L}{2}h^{\mu\lambda}f_{\lambda\nu}
,\qquad \rho^\mu{}_\nu = \tfrac12 h^{\mu\lambda}\un e(h_{\lambda\nu})
,\qquad \varphi_\mu = \un e(a_{\mu})
,\qquad f_{\mu\nu}=D_\mu a_\nu-D_\nu a_\mu
\eeq
and we note that $\varphi_\mu$ is proportional to the  rotation coefficient $C_{e\mu}{}^e$, i.e.,
$\left[\un e,\un D_\mu\right]=L\varphi_\mu \un e$. In addition, we will use the notation\footnote{The notation used here can be interpreted in terms of expansion ($\theta$), shear ($\zeta$), vorticity ($f$) and acceleration ($\varphi$) of the radial congruence $\un e$.} $\theta=tr\rho=\un e (\ln\sqrt{-\det h})$ and $\zeta^\mu{}_\nu=\rho^\mu{}_\nu-\tfrac{1}{d}\theta\delta^\mu{}_\nu$. 
In Appendix \ref{AppA}, we record some additional details, including the Weyl-Riemann curvature components. 

As is the case in the FG gauge, $\gamma^{(0)}_{\mu\nu}(x)$ defines a background boundary metric and acts as a source for the stress energy tensor of the dual field theory, with $\pi^{(0)}_{\mu\nu}(x)$ its vev. We have seen that in WFG gauge, 
$a^{(0)}_\mu(x)$ has the interpretation of a Weyl connection in the dual field theory; in addition, we will show below that $p^{(0)}_\mu(x)$ appears in the Weyl Ward identity as if it were the vev for the Weyl current. We will discuss these operators further in Section \ref{sec:QFT}.

As usual \cite{Skenderis2002}, one finds that the bulk equations of motion determine the subleading component fields in terms of $\gamma^{(0)}_{\mu\nu}(x),a^{(0)}_\mu(x),\pi^{(0)}_{\mu\nu}(x)$ and $p^{(0)}_\mu(x)$. In this paper,  we will confine our attention to vacuum solutions of Einstein gravity that are asymptotically locally anti-de Sitter.
For example, the $ee$-component of the vacuum Einstein equations is
\beq\label{Einee}
0=G_{ee}+\Lambda g_{ee}=-\tfrac12 tr(\rho\rho)-\frac{3L^2}{8} tr(ff)-\tfrac12 \overline{R}+\tfrac12 \theta^2+\Lambda
\eeq
where $\Lambda=-\frac{d(d-1)}{2L^2}$ is the cosmological constant of AdS$_{d+1}$ and we define for the sake of brevity
\beqn\label{barR}
\overline{R}^\lambda{}_{\mu\rho\nu}
=
D_\rho \gamma^\lambda_{\nu\mu}
-D_\nu \gamma^\lambda_{\rho\mu}
+\gamma^\delta_{\nu\mu}\gamma^\lambda_{\rho\delta}
-\gamma^\delta_{\rho\mu}\gamma^\lambda_{\nu\delta}
\eeqn
with $\overline{R}=h^{\mu\nu}\overline{R}^\rho{}_{\mu\rho\nu}$ the corresponding Ricci scalar.
Expanding \eqref{Einee} we find
\beq\label{expandGee}
0 = \left[\Lambda+\frac{d(d-1)}{2L^2}\right]
-\frac12\frac{z^2}{L^2}\left[2(d-1)L^{-2}X^{(1)}+R^{(0)}\right]+...-(d-1)\frac{z^d}{L^d}\left[\frac{d}{2}L^{-2} Y^{(1)}+ \hat\nabla^{(0)}\cdot p_{(0)}\right]+...
\eeq
where $R^{(0)}$ is the boundary Weyl-Ricci scalar and 
\beq
X^{(1)}=\gamma_{(0)}^{\mu\nu}\gamma^{(2)}_{\mu\nu},\qquad
Y^{(1)}=\gamma_{(0)}^{\mu\nu}\pi^{(0)}_{\mu\nu}.\label{trace}
\eeq
In \eqref{expandGee}, the order one equation is trivially satisfied while the $z^2$ contribution gives $X^{(1)}$ entirely in terms of the Weyl-Ricci scalar curvature:
\beq\label{x1}
X^{(1)}=-\dfrac{L^2}{2(d-1)}R^{(0)}.
\eeq
This result looks exactly the same as is obtained in the usual FG calculation, but we stress that the right hand side involves now the Weyl covariant Weyl-Ricci scalar.

For later use, we also note the subleading term in the expansion 
\beq\label{expsqrth}
\sqrt{-\det h(z;x)}= 
\left(\frac{L}{z}\right)^{d}\sqrt{-\det \gamma^{(0)}(x)}\left[1+\frac12\frac{z^2}{L^2}  X^{(1)}+\frac12 \frac{z^4}{L^4}X^{(2)}+...+\frac12\frac{z^d}{L^d} Y^{(1)}+...\right],
\eeq
with $X^{(2)}$ given in \eqref{A19}. Using \eqref{A51} one obtains
\beq
X^{(2)}=-\frac{L^4}{4(d-2)^2}\left[Ric^{(0)}_{\mu\nu}Ric^{(0)\mu\nu}-\frac{d}{4(d-1)} R^{(0)2}-(d-1)tr(f^{(0)2})\right]-\frac{L^2}{2} \hat{\nabla}^{(0)}_\nu a^{(2)\nu}.\label{X2}
\eeq

As in the FG story, we must be careful with the $O(z^d)$ terms here because of divergences in the evaluation of the on-shell action --- those divergences are responsible for the Weyl anomaly in the dual field theory \cite{Henningson1998}, the structure of which we will discuss in detail below. Nevertheless, we may read off the `left-hand-side' of the Weyl Ward identity from eq. \eqref{expandGee},
\beq\label{LHS}
\hat\nabla^{(0)}\cdot p_{(0)}+\dfrac{d} {2L^2}\gamma_{(0)}^{\mu\nu}\pi^{(0)}_{\mu\nu}.
\eeq
We will see later that this is the expected form given the interpretation of $\pi^{(0)}_{\mu\nu}$ and $p^{(0)}_\mu$ in terms of currents in the dual field theory. We will also study the form of the anomalous right-hand-side later, in particular in $d=2$ and $d=4$.

Similarly, one finds that the leading $O(z^2)$ term in $G_{e\mu}$ is proportional to 
\beq\label{Bian}
\gamma_{(0)}^{\lambda\nu}\nabla^{(0)}_\nu\Big(G_{\lambda\mu}^{(0)}+ f^{(0)}_{\lambda\mu}\Big)=0,
\eeq
the vanishing of which is the twice-contracted Bianchi identity of the Weyl connection, as was discussed above (see eq. \eqref{tcBianchi}).

The leading non-trivial terms in the $\mu\nu$-components of the Einstein equations
determine
\beq\label{gamma2}
\gamma_{\mu\nu}^{(2)}=-\frac{L^2}{d-2}\Big( Ric_{(\mu\nu)}^{(0)}
-\frac{1}{2(d-1)}R^{(0)}\gamma_{\mu\nu}^{(0)}\Big)
=-\frac{L^2}{d-2} L^{(0)}_{(\mu\nu)},
\eeq
where $L^{(0)}$ is the Weyl-Schouten tensor. Its trace \eqref{trace} correctly reproduces \eqref{x1}. We take each of these results as representative of the fact that the subleading terms in the expansion of the metric are determined by the Weyl curvature, analogous to what happens in the usual FG gauge in which they are determined by the LC curvature of the induced metric. As we mentioned previously, the difference is that now all of the subleading terms in the bulk fields are Weyl-covariant.

\bigskip

The holographic dictionary for WFG will be taken to be the obvious generalization of the usual relationship \cite{Maldacena:1997re}, i.e.,
\beq
Z_{bulk}[g;\gamma^{(0)}, a^{(0)}]
=exp(-S_{o.s.}[h,a;\gamma^{(0)}, a^{(0)}]) = Z_{FT}[\gamma^{(0)}, a^{(0)}]
\eeq
where on the left we have the on-shell action of the bulk classical theory whose metric is given by $h,a$ with asymptotic configurations $\gamma^{(0)}, a^{(0)}$, while the right-hand-side is the generating functional of correlation functions of operators sourced by $\gamma^{(0)}, a^{(0)}$. Although this is expressed in terms of the `bare' sources, it is implicit that a regularization scheme for the left-hand-side is employed and that the boundary counter-terms are introduced to absorb power divergences that arise in the evaluation of the on-shell action, \cite{Skenderis2002}. Here, we will organize the discussion by taking the space-time dimension $d$ to be formally complex; the on-shell action is convergent for sufficiently small $d$, and as we move $d$ up along the real axis, we encounter additional divergences as $d$ approaches an even integer. It is well-known in the context of Fefferman-Graham that as a byproduct this divergence induces  the Weyl anomaly of the dual field theory, and is associated with the appearance of logarithms in the field expansions when $d$ is precisely an even integer, as discussed in \cite{Henningson1998, Mazur2001}. Here we will review this bit of physics, as the existence of the Weyl connection, as we will see, organizes the Weyl anomaly in a much more symmetric fashion than it is usually described.

It is taken for granted that $Z_{bulk}$ is diffeomorphism invariant. Under the holographic map this implies, among other things, that the dual field theory can be regulated in a diffeomorphism-invariant fashion \cite{Skenderis2002}. However, the bulk calculation is classical, and thus, in principle, is a functional of the bulk metric $g$ as well as the boundary values. We therefore suppose that
\beq
\frac{Z_{bulk}\left[g';\gamma_{(0)}',a_{(0)}',...\Big|z',x'\right]}{Z_{bulk}\left[g;\gamma_{(0)},a_{(0)},...\Big|z,x\right]}=1,
\eeq
where the notation refers to the fact that we are computing the partition function in different coordinate systems. Here of course we are particularly interested in the Weyl diffeomorphism $(z',x')=(z/{\cal B}(x),x)$ which relates the boundary values $\gamma_{(0)}'=\gamma_{(0)}/{\cal B}^2$, $a_{(0)}'=a_{(0)}-\text d\ln {\cal B}$. $Z_{bulk}$ is given in the classical limit by evaluating the (renormalized) on-shell action, $Z_{bulk}=e^{-S_{o.s.}[g;\gamma_{(0)},a_{(0)},...|z,x]}$. We then ask, is it also true that this cleanly induces a Weyl transformation on the boundary?
That is, is it true that
\beq
\frac{Z_{bdy}[x;\gamma'_{(0)},a'_{(0)},...]}{Z_{bdy}[x;\gamma_{(0)},a_{(0)},...]}\stackrel{?}{=}1,
\eeq
where $Z_{bdy}$ is the generating functional in the given background. 
As is well-established (see e.g. \cite{Mazur2001}), what happens is that there is an anomaly 
\beq
\frac{Z_{bulk}\left[g';\gamma_{(0)}',a_{(0)}',...\Big|z',x\right]}{Z_{bulk}\left[g;\gamma_{(0)},a_{(0)},...\Big|z,x\right]}=
e^{-{\cal A}_k}\frac{Z_{bdy}[x;\gamma_{(0)}',a_{(0)}',...]}{Z_{bdy}[x;\gamma_{(0)},a_{(0)},...]}
\eeq
in dimension $d=2k$. Recall that we are employing the specific Weyl diffeomorphism, which is inducing a Weyl transformation on the boundary, but no boundary diffeomorphism. If we take the log of these expressions, the result is that 
\beq\label{anomaly}
0=S_{bulk}[g';\gamma_{(0)}',...|z',x]-S_{bulk}[g;\gamma_{(0)},...|z,x]
=S_{bdy}[x;\gamma_{(0)}',a_{(0)}',...]-S_{bdy}[x;\gamma_{(0)},a_{(0)},...]+{\cal A}_k.
\eeq
That is, when we compare the evaluation of the bulk on-shell action in different coordinate systems, the result appears as the difference of boundary actions in Weyl-equivalent backgrounds, {\it up to an anomalous term, which is not the difference of two such actions}. The only source for such a term is a pole at $d=2k$ in the evaluation of the bulk action, which arises because the on-shell action is not a boundary term, but contains pieces that must be integrated over $z$. The bulk action for Einstein gravity is generally given by ($vol_{S}=\sqrt{-\det h} \ \text d^dx$)
\beq
S_{bulk}[g;\gamma_{(0)},...|z,x]=\dfrac{1}{16\pi G}\int_M e\wedge vol_{S} (R-2\Lambda).
\eeq
On shell, it evaluates to
\beq\label{onshellactionexpansion}
S_{bulk}[g;\gamma_{(0)},...|z,x]=-\dfrac{d}{8\pi G L^2}\int_M e\wedge vol_S=-\dfrac{d}{8\pi G L}\int_{M} \dfrac{\text dz}{z}\wedge \text d^dx \sqrt{-\det h},
\eeq
where we recall that $d$ is the boundary dimension. We then expand $\sqrt{-\det h}$ in powers of $z$, as given in \eqref{expsqrth} to obtain
\beq
S_{bulk}[g;\gamma_{(0)},...|z,x]=-\dfrac{d}{8\pi G L^2}\int_{M} \text dz\wedge \text d^dx \Big(\dfrac{L}{z}\Big)^{d+1}\sqrt{-\det \gamma^{(0)}}\Big(1+\frac12\dfrac{z^2}{L^2}X^{(1)}+\frac12 \frac{z^4}{L^4}X^{(2)}+\dots\Big).
\eeq
Consider now the difference of Weyl-transformed bulk actions as in \eqref{anomaly} and define $vol_\Sigma=\sqrt{-\det\gamma^{(0)}} \ \text d^dx$. The idea is to start with $S_{bulk}[g';\gamma_{(0)}',...|z',x]$, use the explicit Weyl transformation of the different quantities in the expansion (see \eqref{Weyldifftrans}) and then change the name of the integration variable from $z'$ to $z$.\footnote{To evaluate these expressions, a regulator is required. The last step of renaming the integration variable has a corresponding effect on the cutoff and thus is not innocuous in the renormalization procedure. Such a regulator is not Weyl-covariant, which is consistent with the fact that an anomaly arises. Most of the details of the renormalization occur in expressions that are the difference of two Weyl-equivalent actions, whereas the anomaly is not and has been cleanly extracted.} We will demonstrate this for the first two poles, which occur at $d=2$ and $d=4$. Using \eqref{onshellactionexpansion}, we obtain from the left hand side of \eqref{anomaly}
\beqn\label{Wder}
0&=&\dfrac{d}{8\pi GL}\int_M \text d\left(\dfrac{{\cal B}^{-d}}{d}\left(\dfrac{L}{z}\right)^d\right)\wedge vol_\Sigma-\dfrac{d}{8\pi GL}\int_M \text d\left(\dfrac{1}{d}\left(\dfrac{L}{z}\right)^d\right)\wedge vol_\Sigma \nonumber\\
&&+\dfrac{d}{16\pi GL}\int_M \text d\left(\dfrac{{\cal B}^{-(d-2)}}{d-2}\left(\dfrac{L}{z}\right)^{d-2}\right)\wedge {\cal G}^{(1)}_\Sigma-\dfrac{d}{16\pi GL}\int_M\text d\left(\dfrac{1}{d-2}\left(\dfrac{L}{z}\right)^{d-2}\right)\wedge {\cal G}^{(1)}_\Sigma\nonumber\\
&&+\dfrac{d}{16\pi GL}\int_M \text d\left(\dfrac{{\cal B}^{-(d-4)}}{d-4}\left(\dfrac{L}{z}\right)^{d-4}\right)\wedge {\cal G}^{(2)}_\Sigma-\dfrac{d}{16\pi GL}\int_M\text d\left(\dfrac{1}{d-4}\left(\dfrac{L}{z}\right)^{d-4}\right)\wedge {\cal G}^{(2)}_\Sigma+\dots,
\eeqn
with ${\cal G}^{(1)}_\Sigma=X^{(1)} vol_\Sigma$ (Weyl weight $-(d-2)$) and ${\cal G}^{(2)}_\Sigma=X^{(2)} vol_\Sigma$ (Weyl weight $-(d-4)$). 

We now focus our attention to the case $d=2$. We observe that the offending term in $d\to 2^-$ is
\beq\label{A1}
\dfrac{d}{16\pi GL}\int_M \text d\left(\frac{{\cal B}^{-(d-2)}}{d-2}\left(\frac{L}{z}\right)^{d-2}\right)\wedge {\cal G}^{(1)}_\Sigma-\dfrac{d}{16\pi GL}\int_M \text d\left(\frac{1}{d-2}\left(\frac{L}{z}\right)^{d-2}\right)\wedge {\cal G}^{(1)}_\Sigma
=\dfrac{1}{8\pi G L}\int_\Sigma \ln{\cal B}\ {\cal G}^{(1)}_\Sigma.
\eeq
The equality in this equation is obtained expanding ${\cal B}$ around $1$ and eventually imposing $d=2$. For concreteness we rewrite this final result using the holographic value of $X^{(1)}$, \eqref{x1}. Then, we read from \eqref{anomaly}:
\beq
{\cal A}_1=\dfrac{1}{8\pi G L}\int_\Sigma \ln{\cal B}\ {\cal G}^{(1)}_\Sigma=-\dfrac{L}{16\pi G}\int_\Sigma \ln{\cal B} \ R^{(0)}  vol_\Sigma.
\eeq
This numerical coefficient is the correct one, leading to the central charge $c=\dfrac{3L}{2G}$  \cite{brown1986, Henningson1998}. We will shortly comment on the implications, but notice already that $R^{(0)}$ is not the Levi-Civita curvature, as usually found, but rather the Weyl curvature, which depends on both $\gamma^{(0)}$ and $a^{(0)}$. As such, it is a Weyl-covariant scalar.

We move to $d=4$ (the on-shell action itself must be supplemented by  boundary counterterms to move past $d=2$, but these do not contribute to the current computation). Here the pole for $d\to 4^-$ gives
\beq
\dfrac{d}{16\pi GL}\int_M \text d\left(\dfrac{{\cal B}^{-(d-4)}}{d-4}\left(\dfrac{L}{z}\right)^{d-4}\right)\wedge {\cal G}^{(2)}_\Sigma-\dfrac{d}{16\pi GL}\int_M\text d\left(\dfrac{1}{d-4}\left(\dfrac{L}{z}\right)^{d-4}\right)\wedge {\cal G}^{(2)}_\Sigma=\dfrac{1}{4\pi GL}\int_\Sigma \ln{\cal B}\ {\cal G}^{(2)}_\Sigma.
\eeq
Using the result \eqref{X2} for $X^{(2)}$, we explicitly compute
\beq\label{4danom}
{\cal A}_2=-\dfrac{L}{8\pi G}\int_\Sigma \ln{\cal B}\left(L^2\left(\frac18 Ric^{(0)}_{\mu\nu}Ric^{(0)\mu\nu}-\frac{1}{24} R^{(0)2}-\frac{3}{8}tr(f^{(0)2})\right)+\hat{\nabla}^{(0)}_\nu a^{(2)\nu}\right)vol_\Sigma.\label{A2}
\eeq
This result has the same form as familiar expressions given in the  literature \cite{Henningson1998} if $a_\mu$ is set to zero, at which point the last two terms would drop and the first two would then involve the LC curvature of the induced metric (giving the well-known ``$a=c$" result for Einstein gravity).\footnote{In fact, as we will show elsewhere, the anomaly can be written in terms of the Euler character (of the Lorentz connection) and the Weyl tensor squared, so the appearance of $f^{(0)}_{\mu\nu}{}^2$ does not represent a new central charge. We thank Weizhen Jia for pointing this out to us.} There are several differences; first, the curvature tensors appearing in the anomaly as before are Weyl covariant. In addition, we see the appearance of a term of the form $tr(f^{(0)}{}^2)$, where $f^{(0)}_{\mu\nu}$ is the field strength of the Weyl connection $a^{(0)}$. 
We note that in a formalism where $a^{(0)}$ has been set to zero, $f^{(0)}$ is by that assumption flat, but here we are not requiring this. It would be interesting to understand a physical boundary field theory situation where $f^{(0)}$ makes an appearance (note that in holographic fluid states, $f^{(0)}$ is related to vorticity of the fluid state). Finally, notice also that the subleading term $a^{(2)}$ makes an appearance in the anomaly, but only through a total derivative. 

In a general dimension, all of the subleading modes $a_\mu^{(j)}$ will make an appearance. 
It is important to note though that all of the higher modes $a^{(j>0)}_\mu$ are not in fact determined by the equations of motion. This is in keeping with the fact that they represent pure gauge degrees of freedom in the bulk, and here in the $d=4$ Weyl anomaly $a^{(2)}$ makes an appearance as a total derivative ambiguity; as far as the Ward identity is concerned, it can be absorbed into the vev of the Weyl current. One finds that in $d=2k$ dimensions, the mode $a^{(k)}$ appears entirely in just the same way, as a contribution to the anomaly of the form $\hat\nabla^{(0)}\cdot a^{(k)}$. 
Such total derivatives are often simply dropped in discussions of the anomalies, and here we see that they are in fact ambiguities. In higher dimensions, one finds that the modes $a_\mu^{(0<j<k)}$ do not appear in the anomaly --- to the extent that they appear at all, one finds that they multiply constraints such as Bianchi identities satisfied by the background fields. 

As we mentioned previously, the other terms in \eqref{4danom} have the property that if we were to set $a^{(0)}$ to zero the anomaly would reduce to the usual expression involving the Levi-Civita Ricci tensor. One finds that that same property persists in higher dimensions as well (indeed we have computed the $d=6$ and $d=8$ anomalies, the details of which will be presented elsewhere). Of course, if one wants to think entirely from the boundary field theory point of view in which diffeomorphisms and Weyl transformations are background symmetries, then choosing $a^{(0)}=0$ {\it assumes} it is flat, with $f^{(0)}=0$. 

Let us expand on this a little further, by focussing on a simple case in which the boundary metric is conformally flat, $\gamma^{(0)}=e^{2\sigma}\eta$. Given that, one then finds that  the Levi-Civita Ricci tensor is
\beqn
\mathring{R}ic_{(\mu\nu)}=
(d-2)\Big(-\pa_{(\mu}\pa_{\nu)}\sigma+\pa_{\mu}\sigma\pa_{\nu}\sigma\Big)-\Big(\pa^2\sigma-(d-2)(\pa\sigma)^2\Big)\eta_{\mu\nu}
\eeqn
while the Weyl-Ricci curvature is
\beqn
Ric^{(0)}_{(\mu\nu)}
=
(d-2)\Big(\pa_{(\mu}\tilde A_{\nu)}+
\tilde A_{(\mu}\tilde A_{\nu)}
\Big)
+\Big(\pa\cdot \tilde A
-(d-2)\tilde A^2\Big)\eta_{\mu\nu}.
\eeqn
where $\tilde A=a^{(0)}-d\sigma$. So we see that indeed, the Levi-Civita curvature is recovered from the Weyl-Ricci curvature by setting $a^{(0)}$ to zero, while the Weyl-Ricci curvature is Weyl invariant because it depends on the St\"uckelberg-like field $\tilde A$. In the usual formalism, conformally flat metrics generally give a non-zero Ricci curvature that depends on the conformal factor, while in the Weyl connection formalism, the curvature vanishes for all conformally flat metrics {\it if we simultaneously choose $a^{(0)}=d\sigma$ instead of zero}, that is, $\tilde A=0$. The Weyl-Ricci curvature is Weyl-invariant since $\tilde A$ is Weyl invariant. That is, for all data $(\gamma^{(0)}=e^{2\sigma}\eta,a^{(0)}=\pa\sigma)$ in the Weyl orbit of $(\eta,0)$, all Weyl curvatures are zero, while the Levi-Civita curvature is of course zero only for $(\eta,0)$.
Referring to the $d=4$ anomaly  given in \eqref{4danom}, we see that all terms are Weyl invariant. Of course, it is the anomaly coefficients that are of most interest rather than the values that a given curvature polynomial takes in some particular geometry.

As we mentioned, there are terms in the anomaly ($tr f^{(0)}{}^2$ in $d=4$) that have no analogue in the usual formalism (as they are assumed to vanish). It is not clear to us what interpretation there might be for a non-zero $f^{(0)}$ background in a given field theory. We hope to return to this question in a separate publication. It is possible though that Weyl gauge field configurations may play an important role in field theories on space-times of non-trivial topology or with boundaries.

To recap, the Weyl anomaly in $d=2k$ is associated with the difference of two bulk volumes
\beq
\Big(e\wedge {\cal G}^{(k)}_\Sigma\Big)'-\Big(e\wedge {\cal G}^{(k)}_\Sigma\Big)\propto\text d(\ln{\cal B} \ X^{(k)} \ vol_\Sigma),
\eeq
Each term on the left is a closed form (since they are top forms in the bulk), with the difference being an exact form, the exterior derivative of the local Weyl anomaly form.    

The anomalies in the Levi-Civita framework are classifiable and are known explicitly at least through $d=8$ in the usual formalism, and their topological origin apparently understood through BRST methods (see e.g. \cite{Boulanger:2004zf}). We expect that the inclusion of the Weyl connection in the latter story would reorganize it in a useful way, and we expect to return to this in a future publication.  

\section{Field Theory Aspects}\label{sec:QFT}

In this section, we will make some preliminary remarks about the dual field theory, with more detail left for future work.  The holographic analysis implies that we should now consider a field theory coupled to a background metric and Weyl connection, with action $S[\gamma^{(0)},a^{(0)};\Phi]$ where $\Phi$ denotes some collection of dynamical fields to which we will assign definite Weyl weights. As we will explain, this is perfectly natural from the field theory perspective as well, but constitutes a new organization of such field theories (which in the usual formulation are coupled only to a background metric, as thoroughly reviewed for instance in \cite{DHoker:2002nbb}). The quantum theory possesses a partition function $Z[\gamma^{(0)},a^{(0)}]$, obtained by doing the functional integral over the dynamical fields, that depends on the background, both through explicit dependence in the action and in the definition of the functional integral measure. A background Ward identity is generated by changing integration variables $\Phi(x)\mapsto {\cal B}(x)^{w_\Phi}\Phi(x)$ giving\footnote{The formalism can be extended to include sources for any operators, but we restrict our focus here.}
\beq\label{quantumWeylWard}
Z[\gamma^{(0)},a^{(0)}]= e^{-{\cal A}[{\cal B}]} Z[{\cal B}(x)^{-2}\gamma^{(0)},a^{(0)}-\text d\ln {\cal B}(x)]
\eeq
with ${\cal A}$ a possible anomalous contribution. Thus the Weyl Ward identity is a relationship between {\it different} theories, that is, field theories in different backgrounds and so, more properly, we refer to the above equation as a background Ward identity. It is of interest then to consider classical actions that are background Weyl invariant, satisfying $S[\gamma^{(0)},a^{(0)};{\cal B}(x)^{w_\Phi}\Phi]=S[{\cal B}(x)^{-2}\gamma^{(0)},a^{(0)}-\text d\ln {\cal B}(x);\Phi]$. An example is a  free scalar theory, whereby  $w_\Phi=\tfrac12(d-2)$ is the engineering dimension, and the action is
\beq\label{Weylfree}
S[\gamma^{(0)},a^{(0)};\Phi]=-\tfrac12\int \text d^dx\sqrt{-\det \gamma^{(0)}}\ \gamma_{(0)}^{\mu\nu}\hat\nabla^{(0)}_\mu\Phi\hat\nabla^{(0)}_\nu\Phi
\eeq
where $\hat\nabla^{(0)}_\mu\Phi=\pa_\mu\Phi+w_\Phi a_\mu^{(0)}\Phi$ is background Weyl-covariant by itself in the above sense. In the usual formalism without the Weyl connection, the corresponding action is not Weyl invariant, but its Weyl transformation can be cancelled, up to a total derivative, by the addition of a specific additional term proportional to $\int \text d^dx\sqrt{-\det \gamma^{(0)}}\ \mathring{R}\Phi^2$. Here, \eqref{Weylfree} is itself background Weyl invariant, and we may add the independently Weyl invariant term $\int \text d^dx\sqrt{-\det \gamma^{(0)}}\ R^{(0)}\Phi^2$, where $R^{(0)}$ is the Weyl-Ricci scalar, with {\it any} coefficient.  With the presence of the Weyl connection, it is a trivial matter to write a variety of actions with background Weyl invariance.

The background fields, as usual, are interpreted as sources for current operators. For example, the stress tensor of the free theory has the form
\beq\label{92}
\Tc_{\gamma^{(0)},a^{(0)}}^{\mu\nu}(x)=\tfrac{2}{\sqrt{-\det \gamma^{(0)}}}\frac{\delta S[\gamma^{(0)},a^{(0)};\Phi]}{\delta \gamma^{(0)}_{\mu\nu}(x)}
=\hat\nabla_{(0)}^\mu\Phi(x)\hat\nabla_{(0)}^\nu\Phi(x)-\tfrac12 \gamma^{(0)\mu\nu}(x)\gamma^{(0)}{}^{\alpha\beta}(x)\hat\nabla^{(0)}_\alpha\Phi(x)\hat\nabla^{(0)}_\beta\Phi(x).
\eeq
Here we have used pedantic notation to emphasize that the definition of the operator depends on the background fields. This operator is Weyl-covariant, by which we mean
\beq\label{Weylstress}
\Tc_{{\cal B}(x)^{-2}\gamma^{(0)},a^{(0)}-\text d\ln{\cal B}(x)}^{\mu\nu}(x)={\cal B}(x)^{d+2}\Tc_{\gamma^{(0)},a^{(0)}}^{\mu\nu}(x).
\eeq
That is, if we compare correlation functions of the stress tensor in two Weyl-related backgrounds, there will be a relative factor of ${\cal B}(x)^{d+2}$ for each instance of the stress tensor; for brevity, we refer to this as the stress tensor (with two upper indices) having Weyl weight $w_T=d+2$.
Similarly, we have the Weyl current
\beq\label{Weylcurr}
\Jc_{\gamma^{(0)},a^{(0)}}^{\mu}(x)=-\frac{1}{\sqrt{-\det \gamma^{(0)}}}\frac{\delta S[\gamma^{(0)},a^{(0)};\Phi]}{\delta a^{(0)}_{\mu}(x)}
=w_\Phi\Phi(x)\hat\nabla_{(0)}^{\mu}\Phi(x).
\eeqn
This operator is also Weyl-covariant in the same sense as the stress tensor and is of weight $d$. Thus $\Tc^{\mu\nu}$ and $\Jc^{\mu}$ have the properties of  operators that couple to $\gamma^{(0)}_{\mu\nu}$ and $a^{(0)}_\mu$ and appear in the Weyl anomaly in the holographic WFG theory. In a holographic theory, we would not have the free field discussion given here, but we can still discuss sourcing these operators (in a given background). 

Earlier, we saw that the classical Weyl Ward identity involved a linear combination of the trace of the stress tensor and the divergence of the Weyl current. This is in fact easily established in general terms. Here we will use classical language, but the argument easily extends to the quantum case by making use of \eqref{quantumWeylWard}. 
As mentioned above, what we mean by Weyl being a background symmetry is that, classically,
\beq
S[\gamma^{(0)},a^{(0)};{\cal B}^{w_\Phi}\Phi]=S[\gamma^{(0)}/{\cal B}^2,a^{(0)}-\text d\ln{\cal B};\Phi].
\eeq
By expanding both sides for small $\ln{\cal B}$ and going on-shell, we find
\beq
0=-\int \text d^dx \frac{\delta S}{\delta a^{(0)}_\mu(x)}\pa_\mu\ln{\cal B}(x)
+\int \text d^dx \frac{\delta S}{\delta \gamma^{(0)}_{\mu\nu}(x)}\Big(-2\ln{\cal B}(x)\gamma^{(0)}_{\mu\nu}(x)\Big).
\eeq
We recognize that this may be written as
\beq
0=\int \text d^dx\sqrt{-\det\gamma^{(0)}}\ \Jc^\mu(x)\pa_\mu\ln{\cal B}(x)
+\int \text d^dx\sqrt{-\det\gamma^{(0)}}\ \Tc^{\mu\nu}(x)\Big(-\ln{\cal B}(x)\gamma^{(0)}_{\mu\nu}(x)\Big)
\eeq
and, by integrating by parts, we have
\beq\label{QFTWeylWard}
0=-\int \text d^dx\sqrt{-\det\gamma^{(0)}}\Big(\hat{\nabla}^{(0)}_\mu \Jc^\mu(x)+\Tc^{\mu\nu}(x)\gamma^{(0)}_{\mu\nu}(x)\Big)\ln{\cal B}(x).
\eeq
This result serves to identify the relative normalization of $\pi^{(0)}_{\mu\nu}$ and $p^{(0)}_\mu$ (see \eqref{LHS}) and their relation with the currents defined here. Incidentally, the Weyl-covariant derivative appears in \eqref{QFTWeylWard} precisely because the current $\Jc^\mu$ (with raised index) has Weyl weight $d$. 

We remark that typical discussions of related topics are rife with `improvements' to operators such as the stress tensor, including mixing with a so-called `virial current' \cite{Callan1970, DiFrancesco1997}. The operators that we have defined here have the advantage of transforming linearly, and in particular do not mix with each other, under Weyl transformations. Indeed we note the familiar result that, given \eqref{92}, we have
\beq
\gamma^{(0)}_{\mu\nu}\Tc_{\gamma^{(0)},a^{(0)}}^{\mu\nu}(x)
=\frac{2-d}{2}\gamma^{(0)}_{\mu\nu}\hat\nabla_{(0)}^\mu\Phi(x)\hat\nabla_{(0)}^\nu\Phi(x)
\eeq
and thus given \eqref{Weylcurr},
\beq
\hat{\nabla}^{(0)}_\mu \Jc^\mu(x)+\Tc^{\mu\nu}(x)\gamma^{(0)}_{\mu\nu}(x)
=w_\Phi\Phi\hat\nabla_{(0)}^2\Phi+\Big(w_\Phi+\frac{2-d}{2}\Big)\gamma^{(0)}_{\mu\nu}\hat\nabla_{(0)}^\mu\Phi(x)\hat\nabla_{(0)}^\nu\Phi(x)=0
\eeq
So including the Weyl current automatically yields 
the correct (here classical) Weyl Ward identity on-shell. One can interpret the usual improvement of the stress tensor (to be traceless) as the absorption of the Weyl current into a redefinition of the stress tensor.

The added value of our construction is the freedom to source the Weyl current and stress tensor independently, because we interpret the background Weyl connection $a^{(0)}$ as independent from the background metric $\gamma^{(0)}$. In a CFT setting, general diffeomorphisms and Weyl transformations are not symmetries, but instead a conformal diffeomorphism of the boundary metric components can be reabsorbed by a specific Weyl transformation, resulting in a symmetry; that is, the combined transformations give a relationship between backgrounds of the same field theory.  Here, we allow arbitrary Weyl transformations and diffeomorphisms, which relate theories in different backgrounds. Given that in holography we should consider conformal classes of boundary metrics, Weyl transformations should in general be independent from boundary diffeomorphisms. One might  then expect that this boundary Weyl current is physical, with associated non-trivial charges.  
Here we have only touched on a few rudimentary aspects of field theories coupled to Weyl geometry, and will return to further elaboration elsewhere.

\section{Conclusions}\label{conclusion}

In this work, we have discussed the consequences of bringing a Weyl connection into the formulation of holography. In order to address this, we first intrinsically analyzed such connections and their associated geometrical tensors. The need for a Weyl connection arises in theories that, in addition to diffeomorphisms, admit a local rescaling of the metric by an arbitrary local function. The vanishing of the metricity required for the familiar Levi-Civita connection is indeed not maintained under such rescalings, and the Weyl connection is defined as the unique torsionless connection with vanishing Weyl metricity, a Weyl-covariant statement \cite{Folland:1970, Hall:1992}. Although richer than its Levi-Civita counterparts, the geometrical tensors built out of this connection turn out to be quite tractable.

It has long been understood that holographic field theories possess a Weyl invariance, in the sense that they couple not to a metric, but to a conformal class of metrics \cite{Feffe, Fefferman2011}. The introduction of a (background) Weyl connection in holographic field theories is a suitable reformulation in which local Weyl transformations relate such theories in different backgrounds. In our account, the bulk gravitational theory is unmodified, but the gauge-fixing is relaxed (to what we called Weyl-Fefferman-Graham gauge) in such a way that the Weyl diffeomorphisms act geometrically on tensors parametrizing the bulk metric.  The Weyl diffeomorphisms correspond to rescaling the holographic coordinate by functions of the transverse coordinates while leaving the latter unchanged. While the FG expansion induces the LC connection associated to the induced boundary metric \cite{Feffe}, we have proven that the WFG expansion induces on the boundary a Weyl connection. This result indicates that the WFG gauge is the proper bulk parametrization that leaves the bulk diffeomorphisms  corresponding to the boundary Weyl transformations  unfixed. This leads to the interpretation of the Weyl connection in the boundary as a background field together with the boundary metric; essentially, the pair $(\gamma^{(0)},a^{(0)})$ replaces $[\gamma^{(0)}]$. An interesting consequence of the WFG gauge is that the boundary hypersurface is generally not part of a foliation, the distribution that is involved being generally non-integrable. We expect that the details of holographic renormalization require a slightly more sophisticated regulator than is usually employed, but the results of this paper do not rely on such details. 

The WFG gauge involves an expansion in powers of the holographic coordinate in which every coefficient is Weyl-covariant by construction. This result is a powerful reorganization of the holographic dictionary.
The Weyl connection sources a Weyl current which explicitly appears in the subleading expansion of the bulk geometry. Subleading orders of the bulk Einstein equations unravel the boundary Weyl geometrical tensors and relationships between boundary expectation values of the sourced operators. In particular we find the boundary Ward identity relating the trace of the energy-momentum tensor with the divergence of the Weyl current, and in the last section have shown that this is the expected result. 

We then scrutinized the implications of our setup for the Weyl anomaly. Not surprisingly, we found the latter to be given now in terms of Weyl-covariant geometrical objects, instead of the corresponding Levi-Civita objects. We expect that this outcome will have implications for the study and characterization of the anomaly in higher even boundary dimensions. The presence of Weyl geometrical tensors allowed for a cohomological description of the anomaly as a difference of Weyl-related bulk volumes, which offers a clear geometrical interpretation of the anomaly. As mentioned several times in the body of the paper, there are a number of clear followups, particularly on the field theory side, which present themselves, and we look forward to exploring such implications in the future.

\section*{Acknowledgements}

We would like to thank Costas Bachas, Guillaume Bossard, Laurent Freidel, Weizhen Jia, Manthos Karydas, Charles Marteau, Eric Mefford, Tassos Petkou, Marios Petropoulos, Rafael Sorkin and Antony Speranza for interesting discussions related to the results of this paper. This research was supported in part by the US Department of Energy under contract DE-SC0015655, the ANR-16-CE31-0004 contract Black-dS-String and the Perimeter Institute for Theoretical
Physics. Research at Perimeter Institute is supported by the Government of Canada through
the Department of Innovation, Science, and Economic Development Canada and by the
Province of Ontario through the Ministry of Research, Innovation and Science.

\appendix

\section{Details of Bulk Expansions}\label{AppA}
\renewcommand{\theequation}{\thesection.\arabic{equation}}
\setcounter{equation}{0}

We recapitulate here our geometrical setup both in the bulk and in the boundary, and compute the leading orders of the expansion toward $z=0$ of the main quantities involved. These are useful to evaluate Einstein equations order by order, and hence solve for the various geometrical objects. Concretely, we work in the non-coordinate basis 
\beq
\text ds^2=e\otimes e+h_{\mu\nu}\text dx^\mu \otimes \text dx^\nu, \qquad  e=L\Big(\dfrac{\text dz}{z}- a_\mu \text dx^\mu\Big).
\eeq
The dual vectors are
\beq
\un e=L^{-1} z \un \pa_z, \qquad \un D_\mu = \un \pa_\mu+z a_\mu \un \pa_z,
\eeq
and they form an orthonormal basis
\beq
e(\un e)=1, \qquad e(\un D_\mu)=0, \qquad \text dx^\mu(\un D_\nu)=\delta^\mu_{\nu}, \qquad \text dx^\mu(\un e)=0.
\eeq
The vector commutators give
\beq
[\un e,\un D_\mu]=L\un e(a_\mu)\un e=L\varphi_\mu \un e, \qquad [\un D_\mu,\un D_\nu]=L\Big(\un D_\mu a_\nu-\un D_\nu a_\mu\Big)\un e=Lf_{\mu\nu}\un e,
\eeq
from which we read
\beq
C_{e\mu}{}^e=L\varphi_\mu, \qquad C_{\mu\nu}{}^e=Lf_{\mu\nu}, \qquad C_{\mu\nu}{}^\alpha=0.
\eeq
Throughout this Appendix, we refer for brevity to generalized bulk indices as $M=(e,\mu)$ and thus vectors $\un e_M=(\un e,\un D_\mu)$ and metric components $g_{MN}=g(\un e_M,\un e_N)$, the most general non-coordinatized Levi-Civita connection is then
\beq
\Gamma^P_{MN}=\tfrac12 g^{PQ}\Big(\un e_M(g_{NQ})+\un e_N(g_{QM})-\un e_Q(g_{MN})\Big)-\tfrac12 g^{PQ}\Big(C_{MQ}{}^Rg_{RN}+C_{NM}{}^Rg_{RQ}-C_{QN}{}^Rg_{RM}\Big).
\eeq
The metric and its inverse are given in components by
\beq
g_{\mu\nu}=h_{\mu\nu}, \quad g_{e\mu}=0, \quad g_{ee}=1,\quad g^{\mu\nu}=h^{\mu\nu}, \quad g^{\mu e}=0,\quad g^{ee}=1.
\eeq
Then, calling $\theta=tr\rho$ with $\rho^\mu_\nu=\tfrac12 h^{\mu\alpha}\un e(h_{\alpha\nu})$,  the Christoffel symbols evaluate to 
\beqn
\Gamma^e_{ee}&=& 0\\
\Gamma^e_{e\mu}&=& C_{e\mu}{}^e=L\varphi_\mu\\
\Gamma^e_{\mu e}&=& 0\\
\Gamma^e_{\mu\nu}&=& -\tfrac12 \un e(h_{\mu\nu})+\frac{L}{2} f_{\mu\nu}\\
\Gamma^\mu_{ee}&=& h^{\mu\nu}C_{\nu e}{}^e=-Lh^{\mu\nu}\varphi_\nu\\
\Gamma^\mu_{e\nu}&=& \rho^\mu{}_\nu+\frac{L}{2} f^\mu{}_\nu\\
\Gamma^\mu_{\nu e}&=& \rho^\mu{}_\nu+\frac{L}{2} f^\mu{}_\nu\\
\Gamma^\mu_{\mu e}&=&\theta\\
\Gamma^\mu_{\alpha\beta}&=& \tfrac12 h^{\mu\nu}\Big( \un D_\alpha h_{\beta\nu}+\un D_\beta h_{\alpha \nu}-\un D_\nu h_{\alpha\beta}\Big)\equiv \gamma^\mu_{\alpha\beta}.
\eeqn
These connections are explicitly reported in \eqref{conn1}, \eqref{conn2}, \eqref{conn3} and \eqref{conn4}.
We additionally define
\beq
m_{(k)}{}^\mu{}_\nu\equiv (\gamma_{(0)}^{-1}\gamma_{(k)})^\mu{}_\nu,\qquad
n_{(k)}{}^\mu{}_\nu\equiv (\gamma_{(0)}^{-1}\pi_{(k)})^\mu{}_\nu,
\eeq
and the scalars \beqn
X^{(1)}&=&tr(m_{(2)}),\\
X^{(2)}&=&tr( m_{(4)})
-\tfrac12 tr( m_{(2)}^2)
+\tfrac14 \Big( tr (m_{(2)})\Big)^2,\label{A19}\\
Y^{(1)}&=&tr (n_{(0)}).
\eeqn
Starting from the metric \eqref{FGexpandmetricBetter} and the Weyl connection \eqref{WFGexpandgauge} expansions, we compute the inverse metric, the determinant and the various connection components appearing in \eqref{connComp}. We expand the two series enough to be able to capture the two leading orders. The result is:
\beqn
h^{\mu\lambda}(z;x)&=&\frac{z^2}{L^2}\left[\gamma_{(0)}^{-1}
-\frac{z^2}{L^2}m_{(2)}\gamma_{(0)}^{-1}
-\frac{z^4}{L^4}(m_{(4)}-m_{(2)}^2)\gamma_{(0)}^{-1}+...\right]^{\mu\lambda}
-\frac{z^{d+2}}{L^{d+2}}\left[n_{(0)}\gamma_{(0)}^{-1}+...\right]^{\mu\lambda}
\\\label{metricdet}
\sqrt{-\det h(z;x)}&=& 
\left(\frac{L}{z}\right)^{d}\sqrt{-\det \gamma^{(0)}(x)}\left[1+\frac12\frac{z^2}{L^2}  X^{(1)}+\frac12 \frac{z^4}{L^4}X^{(2)}+...+\frac12\frac{z^d}{L^d} Y^{(1)}+...\right]
\\
\rho^\mu{}_\nu(z;x)&=&L^{-1}\left[-\delta^\mu{}_\nu
+\frac{z^2}{L^2}m_{(2)}{}^\mu{}_\nu
+\frac{z^4}{L^4}(2m_{(4)}-m_{(2)}^2)^\mu{}_\nu+...
+\frac{d}{2}\frac{z^d}{L^d}n_{(0)}{}^\mu{}_\nu+...\right]
\\
\theta(z;x)&=&L^{-1}\left[-d
+\frac{z^2}{L^2}X^{(1)}+\dfrac{z^4}{L^4}2(X^{(2)}-\frac{1}{4}(X^{(1)})^2)+...
+\frac{d}{2}\frac{z^d}{L^d}Y^{(1)}+...
\right]
\\
\varphi_\mu(z;x)&=&L^{-1}\left[ \frac{z^2}{L^2}2a_\mu^{(2)}+...
+\frac{z^{d-2}}{L^{d-2}}(d-2)p_\mu^{(0)}+...\right]
\\
f_{\mu\nu}(z;x)&=&f^{(0)}_{\mu\nu}(x)+\frac{z^2}{L^2}(\hat\nabla^{(0)}_\mu a^{(2)}_{\nu}-\hat\nabla^{(0)}_\nu a^{(2)}_{\mu})+...
+\frac{z^{d-2}}{L^{d-2}}(\hat\nabla^{(0)}_\mu p^{(0)}_{\nu}-\hat\nabla^{(0)}_\nu p^{(0)}_{\mu})+...
\eeqn
with $f^{(0)}_{\mu\nu}=\pa_\mu a^{(0)}_\nu-\pa_\nu a^{(0)}_\mu$. In the expression for $f_{\mu\nu}$ we used the boundary derivative introduced in \eqref{WeylGauged}, which is the Weyl derivative shifted with the Weyl weight of the object it acts upon. For instance, looking at \eqref{Weyldifftrans2}, $a^{(2)}_\mu$ and $p^{(0)}_\mu$ are Weyl-covariant with weights $2$ and $d-2$ respectively and therefore:
\beqn
\hat \nabla^{(0)}_\mu a^{(2)}_\nu &=& \nabla^{(0)}_\mu a^{(2)}_\nu+2 a^{(0)}_\mu a^{(2)}_\nu,\\
\hat \nabla^{(0)}_\mu p^{(0)}_\nu &=& \nabla^{(0)}_\mu p^{(0)}_\nu+(d-2) a^{(0)}_\mu p^{(0)}_\nu,
\eeqn
with $\nabla^{(0)}$ the boundary Weyl connection (its connection coefficients are explicitly given in \eqref{InducedWeyl}). 

The expansion of the geometrical objects constructed from \eqref{barR} is also reported
\beqn
\gamma_{\mu\nu}^\lambda
&=&
\gamma^{(0)}_{\mu\nu}{}^\lambda
+\frac{z^{2}}{L^{2}}\Big[
\tfrac12\gamma_{(0)}^{\lambda\xi}\Big(\hat\nabla^{(0)}_\nu \gamma^{(2)}_{\mu\xi}
+\hat\nabla^{(0)}_\mu \gamma^{(2)}_{\xi\nu}
-\hat\nabla^{(0)}_\xi \gamma^{(2)}_{\mu\nu}\Big)
-\Big(
a_\mu^{(2)}\delta^{\lambda}{}_{\nu}
+a_\nu^{(2)}\delta^{\lambda}{}_{\mu}
-a_\xi^{(2)}\gamma_{(0)}^{\lambda\xi}\gamma_{\mu\nu}^{(0)}\Big)
\Big]+...
\nonumber\\&&
-\frac{z^{d-2}}{L^{d-2}}\left[
p^{(0)}_{\mu}\delta^{\lambda}{}_{\nu}
+p^{(0)}_{\nu}\delta^{\lambda}{}_{\mu}
-p^{(0)}_{\rho}\gamma_{(0)}^{\lambda\rho}\gamma^{(0)}_{\mu\nu}\right]+...
\\
\overline{Ric}_{\mu\nu}
&=&
{Ric}^{(0)}_{\mu\nu}+\frac{z^{2}}{L^{2}}\Big[
\tfrac12\hat\nabla^{(0)}_\lambda\Big(\gamma_{(0)}^{\lambda\xi}\Big(\hat\nabla^{(0)}_\nu \gamma^{(2)}_{\mu\xi}
+\hat\nabla^{(0)}_\mu \gamma^{(2)}_{\xi\nu}
-\hat\nabla^{(0)}_\xi \gamma^{(2)}_{\mu\nu}\Big)\Big)
\nonumber\\&&+(d-1)\hat\nabla^{(0)}_\nu a_\mu^{(2)}
-\hat\nabla^{(0)}_\mu a_\nu^{(2)}
+\gamma_{\mu\nu}^{(0)}\hat\nabla^{(0)}\cdot a^{(2)}
-\tfrac12\hat\nabla^{(0)}_\nu\hat\nabla^{(0)}_\mu X^{(1)}
\Big]
\nonumber\\&&+...
+\frac{z^{d-2}}{L^{d-2}}
\Big[(d-1)\hat\nabla^{(0)}_\nu p_\mu^{(0)}
-\hat\nabla^{(0)}_\mu p_\nu^{(0)}
+\gamma^{(0)}_{\mu\nu}\hat\nabla^{(0)}\cdot p^{(0)}
\Big]+...\\
\overline{R}&=&\frac{z^{2}}{L^{2}}R^{(0)}
+\frac{z^{4}}{L^{4}}\Big[
\gamma_{(0)}^{\lambda\nu}\hat\nabla^{(0)}_\lambda\hat\nabla^{(0)}_\mu \Big(m_{(2)}{}^{\mu}{}_{\nu}-tr (m_{(2)})\delta^\mu{}_\nu\Big)
+2(d-1)\hat\nabla^{(0)}\cdot a^{(2)}
-tr(m_{(2)}\gamma_{(0)}^{-1}{Ric}^{(0)})\Big]
\nonumber\\&&+...
+2(d-1)\frac{z^{d}}{L^{d}}
\hat\nabla^{(0)}\cdot p^{(0)}+...
\\
\overline{G}_{\mu\nu}&=&{G}^{(0)}_{\mu\nu}
+\frac{z^2}{L^2}\Big[\tfrac12\hat\nabla^{(0)}_\lambda\Big(\gamma_{(0)}^{\lambda\xi}\Big(\hat\nabla^{(0)}_\nu \gamma^{(2)}_{\xi\mu}
+\hat\nabla^{(0)}_\mu \gamma^{(2)}_{\xi\nu}
-\hat\nabla^{(0)}_\xi \gamma^{(2)}_{\mu\nu}\Big)\Big)
+(d-1)\hat\nabla^{(0)}_\nu a_\mu^{(2)}
-\hat\nabla^{(0)}_\mu a_\nu^{(2)}\nonumber\\&&
-(d-2)\gamma_{\mu\nu}^{(0)}\hat\nabla^{(0)}\cdot a^{(2)}
-\tfrac12\hat\nabla^{(0)}_\nu\hat\nabla^{(0)}_\mu X^{(1)}
-\tfrac12\gamma^{(2)}_{\mu\nu}R^{(0)}
-\tfrac12\gamma^{(0)}_{\mu\nu}\hat\nabla^{(0)}_\lambda\hat\nabla^{(0)}_\phi \Big((\gamma_{(0)}^{-1}\gamma^{(2)}\gamma_{(0)}^{-1})^{\phi\lambda}-X^{(1)}\gamma_{(0)}^{\phi\lambda}\Big)\nonumber\\&&
+\tfrac12\gamma^{(0)}_{\mu\nu} tr(m_{(2)}\gamma_{(0)}^{-1}{Ric}^{(0)})\Big]+...\nonumber\\&&
+\frac{z^{d-2}}{L^{d-2}}
\Big[(d-1)\hat\nabla^{(0)}_\nu p_\mu^{(0)}
-\hat\nabla^{(0)}_\mu p_\nu^{(0)}
-(d-2)\hat\nabla^{(0)}\cdot p^{(0)}\gamma^{(0)}_{\mu\nu}
\Big]+...
\eeqn
These quantities appear explicitly in the Einstein tensor. We then compute the bulk Ricci tensor:
\beq
Ric_{MN}=R^P{}_{MPN}=\un e_P (\Gamma^P_{NM})-\un e_N(\Gamma^P_{PM})+\Gamma^Q_{NM}\Gamma^P_{PQ}-\Gamma^Q_{PM}\Gamma^P_{NQ}-C_{PN}{}^Q\Gamma^P_{QM},
\eeq
and so
\beqn
Ric_{ee}&=& -L\nabla_\mu \varphi^\mu-L^2\varphi^2-\un e(\theta)-tr(\rho\rho)-\frac{L^2}{4} tr(ff)\\
Ric_{e\mu} &=& \nabla_\alpha\Big(\rho^\alpha_\mu+\frac{L}{2} f^\alpha{}_\mu\Big)-\un D_\mu \theta+L^2\varphi^\alpha f_{\alpha\mu}\\
Ric_{\mu e}&=& \nabla_\alpha\Big(\rho^\alpha_\mu+\frac{L}{2} f^\alpha{}_\mu\Big)-\un D_\mu \theta+L^2\varphi^\alpha f_{\alpha\mu}\\
Ric_{\mu\nu}&=& \overline{Ric}_{\mu\nu}-L\nabla_\nu \varphi_\mu-(\un e+\theta)\Big(\rho_{\mu\nu}+\frac{L}{2} f_{\mu\nu}\Big)-L^2\varphi_\mu\varphi_\nu+2\rho_\mu^\alpha\rho_{\alpha\nu}+\frac{L^2}{2} f_{\nu\alpha}f^\alpha{}_\mu.
\eeqn
Notice that $Ric_{e\mu}=Ric_{\mu e}$. The trace of the Ricci tensor gives the scalar curvature
\beq
R = g^{MN}\Big(\un e_P (\Gamma^P_{NM})-\un e_N(\Gamma^P_{PM})+\Gamma^Q_{NM}\Gamma^P_{PQ}-\Gamma^Q_{PM}\Gamma^P_{NQ}-C_{PN}{}^Q\Gamma^P_{QM}\Big).
\eeq
It evaluates to
\beq
R=-2 \un e(\theta)+\frac{L^2}{4} tr(ff)-tr(\rho\rho)-2Lh^{\mu\nu}\nabla_\mu \varphi_\nu+\overline{R}-\theta^2-2L^2\varphi_\mu \varphi_\nu h^{\mu\nu}.
\eeq
Therefore the various components of the Einstein tensor read
\beqn
G_{ee}&=&-\tfrac12 tr(\rho\rho)-\frac{3L^2}{8} tr(ff)-\tfrac12 \overline{R}+\tfrac12 \theta^2\\
G_{e\mu}&=& \nabla_\alpha\Big(\rho^\alpha_\mu+\frac{L}{2} f^\alpha{}_\mu\Big)-\un D_\mu \theta+L^2\varphi^\alpha f_{\alpha\mu}\\
G_{\mu e}&=& \nabla_\alpha\Big(\rho^\alpha_\mu+\frac{L}{2} f^\alpha{}_\mu\Big)-\un D_\mu \theta+L^2\varphi^\alpha f_{\alpha\mu}\\
G_{\mu\nu}&=& \overline{G}_{\mu\nu}-L\nabla_\nu \varphi_\mu-(\un e+\theta)\Big(\rho_{\mu\nu}+\frac{L}{2} f_{\mu\nu}\Big)-L^2\varphi_\mu\varphi_\nu+2\rho_\mu^\alpha\rho_{\alpha\nu}+\frac{L^2}{2} f_{\nu\alpha}f^\alpha{}_\mu \\
&& +h_{\mu\nu}\Big(\un e(\theta)-\frac{L^2}{8} tr(ff)+\tfrac12 tr(\rho\rho)+L\nabla_\alpha \varphi^\alpha+\tfrac12\theta^2+L^2\varphi^2\Big).
\eeqn
Finally, vacuum Einstein equations are given by
\beq
G_{MN}+\Lambda g_{MN}=0.
\eeq
They become
\beqn\label{ee}
0&=&-\tfrac12 tr(\rho\rho)-\frac{3L^2}{8} tr(ff)-\tfrac12 \overline{R}+\tfrac12 \theta^2+\Lambda\\\label{emu}
0&=& \nabla_\alpha\Big(\rho^\alpha_\mu+\frac{L}{2} f^\alpha{}_\mu\Big)-\un D_\mu \theta+L^2\varphi^\alpha f_{\alpha\mu}\\
0&=& \nabla_\alpha\Big(\rho^\alpha_\mu+\frac{L}{2} f^\alpha{}_\mu\Big)-\un D_\mu \theta+L^2\varphi^\alpha f_{\alpha\mu}\\\label{munu}
0&=& \overline{G}_{\mu\nu}-L\nabla_\nu \varphi_\mu-(\un e+\theta)\Big(\rho_{\mu\nu}+\frac{L}{2} f_{\mu\nu}\Big)-L^2\varphi_\mu\varphi_\nu+2\rho_\mu^\alpha\rho_{\alpha\nu}+\frac{L^2}{2} f_{\nu\alpha}f^\alpha{}_\mu \\
&& +h_{\mu\nu}\Big(\un e(\theta)-\frac{L^2}{8} tr(ff)+\tfrac12 tr(\rho\rho)+L\nabla_\alpha \varphi^\alpha+\tfrac12\theta^2+L^2\varphi^2+\Lambda\Big).
\eeqn
We can obtain relationships among all the various terms in the expansion of $h_{\mu\nu}$ and $a_\mu$ by solving these equations order by order in $z$. For instance, \eqref{ee} is expanded in \eqref{expandGee} whereas the expansion of \eqref{emu} gives \eqref{Bian}. Eventually, expanding \eqref{munu} we obtain at first non-trivial order $\gamma^{(2)}_{\mu\nu}$ as written in \eqref{gamma2}. There exists a particular combination of the previous equations which is simply solved as it does not involve curvature terms:
\beq
0=\frac{g^{MN}(G_{MN}+\Lambda g_{MN})}{d-1}-(G_{ee}+\Lambda g_{ee})=\un e(\theta)+L\nabla_\nu \varphi^\nu+L^2\varphi^2+tr\psi^2-\frac{d}{L}. \label{A51}
\eeq
This was used for example in \eqref{X2} in the process of obtaining $X^{(2)}$.
\bibliographystyle{uiuchept}
\bibliography{adsflatbib}
\end{document}